\documentclass{amsart}
\usepackage{amsmath}
\usepackage{graphicx}
\usepackage{amsthm}
\usepackage{bm}

\newcommand{\nc}{\newcommand}
\nc{\R}{\mathbb{R}}
\nc{\bx}{{\bf x}}
\nc{\by}{{\bf y}}
\nc{\Bs}{{\bf s}}
\nc{\bd}{{\bf d}}
\nc{\bF}{{\bf F}}
\nc{\bFc}{\bm{\mathcal F}}
\nc{\bEc}{\bm{\mathcal E}}
\nc{\bG}{{\bf G}}
\nc{\bone}{\bm{1}}
\nc{\s}{\sigma}
\nc{\bg}{{\bm{\gamma}}}
\nc{\bss}{{\bm{\sigma}}}
\nc{\be}{{\bm{\epsilon}}}
\nc{\bka}{{\bm{\kappa}}}
\nc{\bkappa}{{\bm{\kappa}}}
\nc{\boeta}{{\bm{\eta}}}
\nc{\bpi}{{\bm{\pi}}}
\nc{\B}{\mathcal B}
\nc{\I}{\mathcal I}
\nc{\G}{\Gamma}
\nc{\Pc}{\mathcal P}
\nc{\E}{\mathcal E}
\nc{\U}{\mathcal U}
\nc{\Rm}{\mathcal R}
\nc{\M}{\mathcal M}
\nc{\Q}{\mathcal Q}
\nc{\Sc}{\mathcal S}
\nc{\D}{\mathcal D}
\nc{\K}{\mathcal K}
\nc{\F}{\mathcal F}
\nc{\Gc}{\mathcal G}
\nc{\C}{\mathcal C}
\nc{\ppr}{\bpi_{_{\mathrm{pr}}}}
\nc{\sM}{\s_{_{\mathrm{MAP}}}}
\nc{\BsM}{\Bs_{_{\mathrm{MAP}}}}
\nc{\os}{\overline{\sigma}}
\nc{\reff}{{\mathrm{ref}}}

\nc{\bbF}{\mathbb{F}}
\nc{\real}{\mathbb{R}}
\nc{\bbQ}{\mathbb{Q}}

\newtheorem{remark}{Remark}

\newcommand{\rlab}[1]{\parbox[b][0.3\textwidth][c]{1em}{\rotatebox{90}{#1}}}

\title[Noise effects on EIT with resistor networks]
{Study of noise effects in electrical impedance tomography with
resistor networks}

\author[L. Borcea, F. Guevara Vasquez and A.V. Mamonov]{}

         
\email{borcea@rice.edu}
\email{fguevara@math.utah.edu}
\email{mamonov@ices.utexas.edu}

\begin{document}
\maketitle

\centerline{\scshape Liliana Borcea}
\medskip
{\footnotesize
 \centerline{Computational and Applied Mathematics, Rice University,}
 \centerline{MS 134, 6100 Main St. Houston, TX 77005-1892, USA}
} 

\medskip

\centerline{\scshape Fernando Guevara Vasquez}
\medskip
{\footnotesize
 \centerline{Department of Mathematics, University of Utah,}
 \centerline{155 S 1400 E RM 233, Salt Lake City, UT 84112-0090, USA}
} 

\medskip

\centerline{\scshape Alexander V. Mamonov}
\medskip
{\footnotesize
 \centerline{Institute for Computational Engineering and Sciences, University of
  Texas at Austin,}
  \centerline{1 University Station C0200, Austin, TX 78712, USA}
} 

\bigskip

\begin{abstract}
We present a study of the numerical solution of the two dimensional
electrical impedance tomography problem, with noisy measurements of
the Dirichlet to Neumann map. The inversion uses parametrizations of
the conductivity on optimal grids.  The grids are optimal in the sense
that finite volume discretizations on them give spectrally accurate
approximations of the Dirichlet to Neumann map. The approximations are
Dirichlet to Neumann maps of special resistor networks, that are
uniquely recoverable from the measurements.  Inversion on optimal
grids has been proposed and analyzed recently, but the study of noise
effects on the inversion has not been carried out.  In this paper we
present a numerical study of both the linearized and the nonlinear
inverse problem. We take three different parametrizations of the
unknown conductivity, with the same number of degrees of freedom. We
obtain that the parametrization induced by the inversion on optimal
grids is the most efficient of the three, because it gives the
smallest standard deviation of the maximum a posteriori estimates of
the conductivity, uniformly in the domain. For the nonlinear problem
we compute the mean and variance of the maximum a posteriori estimates
of the conductivity, on optimal grids. For small noise, we obtain that
the estimates are unbiased and their variance is very close to the
optimal one, given by the Cram\'{e}r-Rao bound. For larger noise we
use regularization and quantify the trade-off between reducing the
variance and introducing bias in the solution. Both the full and
partial measurement setups are considered.
\end{abstract}

\section{Introduction \label{sec:intro}}
We study the inverse problem of electrical impedance tomography (EIT)
in two dimensions, with noisy measurements of the Dirichlet to Neumann
(DtN) map. Explicitly, we seek the positive and bounded, scalar valued
coefficient $\s(\bx)$ in the elliptic equation
\begin{equation}
\nabla \cdot \left[ \s(\bx) \nabla u(\bx) \right] = 0, \qquad \bx \in
\Omega.
\label{eq:EIT1}
\end{equation} 
The domain $\Omega$ is bounded, simply connected, with smooth
boundary $\B$. By the Riemann mapping theorem all such domains in $\R^2$ are
conformally equivalent, so from now on we take for $\Omega$ the unit
disk.  We call $\sigma(\bx)$ the conductivity and $u \in
H^1(\Omega)$ the potential, satisfying the boundary conditions
\begin{equation}
u(\bx) = V(\bx), \qquad \bx \in \B,
\label{eq:EIT2}
\end{equation}
for arbitrary $V \in H^{1/2}(\B)$.  The data are finitely many noisy
measurements of the DtN map $\Lambda_\s:H^{1/2}(\B) \to H^{-1/2}(\B)$,
which takes the boundary potential $V$ to the normal boundary flux
(current)
\begin{equation}
\Lambda_\s V(\bx) = \s(\bx) \frac{\partial u({\bf x})}{\partial n},
\qquad \bx \in \B.
\label{eq:EIT3}
\end{equation}
We consider both the \emph{full boundary} setup, where $\Lambda_\s V$
is measured all around the boundary $\B$, and the \emph{partial
  boundary} setup, where the measurements are confined to an
accessible subset $\B_A \subset \B$, and the remainder $\B_I = \B
\setminus \B_A$ of the boundary is assumed grounded $(V|_{\B_I} = 0)$.

In theory, full knowledge of the DtN map $\Lambda_\s$ determines
uniquely $\s$, as proved in \cite{nachman1996gut,brown1997uniqueness}
under some smoothness assumptions on $\sigma$, and in
\cite{astala2005cip} for bounded $\sigma$. The result extends to the
partial boundary setup, at least for $\s \in C^{3 + \epsilon} ( \bar
\Omega )$, $\epsilon > 0$ as established in
\cite{imanuvilov2008gup}. In practice, the difficulty lies in the
exponential instability of EIT. It is shown in
\cite{alessandrini1988sdc,barcelo2001sic,mandache2001eii} that the
best possible stability estimate is of logarithmic type.  Thus even if
the noisy data is consistent (i.e. in the set of DtN maps) we need
exponentially small noise to get a conductivity that is close to the
true one.

It is shown in \cite{alessandrini2005lipschitz} that if $\sigma$ has
finitely many degrees of freedom, more precisely if it is piecewise
constant with a bounded number of unknown values, then the stability
estimates on $\s$ are of Lipschitz type. However, it is not clear how
the Lipschitz constant grows depending on the distribution of the
unknowns in $\Omega$. For example, it should be much easier to
determine the value of $\sigma$ near the boundary than in a small set
in the interior of $\Omega$. 

An important question is how to find 
parametrizations of $\s$ that capture the trade-off between stability
and resolution as we move away from the boundary, where the
measurements are made.  On one hand, the parametrizations should be
sparse, with a small number of degrees of freedom.  On the other hand,
the parametrizations should be adaptively refined toward the boundary.

Adaptive parametrizations for EIT have been proposed in
\cite{isaacson,macmillan2004first} and in
\cite{ameur2002refinement,ameur2002regularization}.  The first
approaches use \emph{distinguishability grids} that are defined with a
linearization argument. The approach in
\cite{ameur2002refinement,ameur2002regularization} is nonlinear and
consists of an iterative coarsening and refinement of a piecewise
constant discretization of the conductivity, with each
discretization update being computationally costly.

We follow the approach in
\cite{BorDruGue,seagar1983probing,BDM-10,BDMG-10} and parametrize $\s$
on \emph{optimal grids}.  The number of parameters is limited by the
noise level in the measurements and their geometrical distribution in
$\Omega$ is determined as part of the inversion. The grids are based
on rational approximations of the DtN map. We call them optimal
because they give spectral accuracy of approximations of $\Lambda_\s$
with finite volume schemes.  The grids turn out to be refined near the
accessible boundary, where we make the measurements, and coarse
away from it, thus capturing the expected loss of resolution of the
reconstructions of $\s$.

Optimal grids were introduced in \cite{asvadurov2000adg, DruKni,
druskin2000gsr, IngDruKni} for accurate approximations of the DtN map in
forward problems. Inversion on optimal grids was first proposed  for
Sturm-Liouville inverse spectral problems in \cite{BorDru}. The analysis
in \cite{BorDruKni} shows that optimal grids are necessary and
sufficient for convergence of solutions of discrete inverse
spectral problems to the true solution of the continuum one. The
numerical solution of EIT on optimal grids was introduced in
\cite{BorDruGue, GuevaraPhD} for the full boundary measurements case,
and in \cite{BDM-10,BDMG-10,MamonovPhD} for partial boundary
measurements.  The inversion in \cite{BorDruGue,
GuevaraPhD,BDM-10,BDMG-10,MamonovPhD} is based on the rigorous theory of
discrete inverse problems for circular planar resistor networks
\cite{CurtMooMor,CurtIngMor,IngerLayer,deverdiere1994rep,
deverdiere1996rep}, which gives networks that can be uniquely determined
by discrete measurements of the continuum DtN map
\cite{MorInger,BorDruGue}. Just as in the continuum EIT, the inverse
problem for networks is ill-posed, and there is a trade-off between the
size of the network and the stability of the reconstruction. 

We present here a study of the inversion algorithms on optimal grids,
for noisy measurements of the DtN map.  We fix the number $g$ of
degrees of freedom, and analyze the effect of the adaptive
parametrization of $\s$ on the reconstruction error.  We consider
maximum a posteriori estimates of $\s$ for the linearized problem about
a constant conductivity, and for the nonlinear problem. The noise is
mean zero Gaussian, and if its standard deviation is small, the only
prior on $\s$ is that it is positive and bounded. For larger noise we
use regularization (Gaussian priors), and study how the
parametrization affects the trade-off between the stability of the
result and the bias.

We study three different parametrizations of $\s$, with $g$ degrees of
freedom. The first two are piecewise linear, on an equidistant grid and
on the optimal grid. The only relation between the second
parametrization and resistor network inversion on optimal grids is the
location of the grid nodes. The third parametrization is that induced by
the resistor network inversion.

In the linearization study we compute the standard deviation of the
estimates and show that the resistor network parametrization is clearly
superior. It gives estimates with uniformly small standard deviation in
$\Omega$. The conclusion is that it is not enough to distribute the
parameters on the optimal grid to obtain good results. To control the
stability of the reconstructions, we also need to use proper basis
functions.

In the nonlinear statistical study we compute maximum a posteriori
estimates of $\s$ with the inversion algorithms on optimal grids.  We
assess their quality by displaying pointwise in $\Omega$ their mean
and standard deviation.  We obtain that the resistor network based
inversion is efficient in the sense that it gives unbiased estimates
of $\s$, with variance that is very close to the optimal
Cram\'{e}r-Rao bound \cite{schervish1995theory}.  This is for small
noise. For larger noise we use regularization priors that introduce
some bias in the solution.  We also compare the network based
inversion to the usual optimization approach that seeks the
conductivity as the least squares minimizer of the data misfit.  For
the optimization, the conductivity is piecewise linear with the same
number $g$ of degrees of freedom, either on a uniform grid or on the
optimal grid. Our numerical experiments indicate that for a fixed
allowed error (standard deviation) in the reconstructions, the network
based method gives reconstructions that are closer in average to the
true conductivity (i.e. with less bias). The conclusion for the
non-linear problem is similar to that for the linearized problem: the
reconstruction error is reduced with the network based inversion as
compared to optimization on either equidistant or optimal grids.  Our
study considers both the full and partial measurement setups
\cite{BorDruGue, GuevaraPhD,BDM-10,BDMG-10,MamonovPhD}.

The paper is organized as follows: We begin in
section~\ref{sect:Bayesian} with the estimation framework. Then we
review the resistor network inversion in section~\ref{sect:Rnets}.
The tools needed for the numerical experiments are described in
section~\ref{sect:setup}.  The estimation results are in section
\ref{sect:NumericalRes}. We end with a summary in section
\ref{sect:Summary}.

\section{Maximum a posteriori estimation of the conductivity}
\label{sect:Bayesian}
\setcounter{equation}{0} 

We study how different parametrizations of the unknown conductivity,
with a fixed number $g$ of degrees of freedom, affect the sensitivity
of the reconstructions to noise in the data. Let $\Bs = (s_1, \ldots,
s_g)^T \in \R^g$ be the vector of parameters, and  
\begin{equation}
\s(\bx) = \left[\Sc(\Bs)\right](\bx), \quad \bx \in \bar{\Omega}
\label{eq:interp}
\end{equation}
the parametrization of the reconstruction conductivity, using an
operator $\Sc:\mathbb{R}^g \to C(\Omega)$ that takes $\Bs$ to a
continuous function in $\Omega$. Since the data is noisy, the
reconstructions are random variables.  We study maximum a posteriori
estimates of the reconstructions under certain priors, as explained in
section~\ref{sect:Bayespdf}. We consider three different
parametrizations of the form (\ref{eq:interp}), outlined in
section~\ref{sect:param}.

\subsection{\textbf{Estimation}}
\label{sect:Bayespdf}
We denote by $\bF: C(\Omega) \to \mathbb{R}^{g}$, the forward map that
associates to a conductivity $\sigma$ the vector of measurements
$\bF(\sigma)$ of the DtN map. The measurement operation is explained
in detail in section~\ref{sect:meas}. It amounts to recording 
voltages and currents at $n$ electrodes on the boundary. The dimension
$g=n(n-1)/2$ of the data vector $\bd$ corresponds to the number of
independent measurements that can be made with $n$ electrodes. The
data model is 
\begin{equation}
\bd = 
\bF(\sigma) + \be, \qquad \be \in
     {\mathcal N} \left({\bf 0},\C\right),
\label{eq:data}
\end{equation}
with $\be$ the noise vector.  The notation ${\mathcal N} \left({\bf
0},\C\right)$ states that $\be$ is Gaussian (multivariate normal),
with mean zero and diagonal covariance $\C$. We refer 
to section~\ref{sect:meas} for an explanation of the uncorrelation
of the components of $\be$.

We parametrize the conductivity as in (\ref{eq:interp}) with a vector
$\Bs \in \mathbb{R}^g$, with $g=n(n-1)/2$ being the same dimension as
the data space. Since the data $\bd$ is tainted with noise, we treat
$\Bs$ as a continuum random variable, and denote by $\ppr(\Bs)$ its
prior probability density. The likelihood function $\bpi(\bd|\Bs)$ is
the probability density of $\bd$ conditioned to knowing $\Bs$. Given
our Gaussian noise model, it takes the form
\begin{equation}
\bpi(\bd|\Bs) = \frac{1}{(2 \pi)^{g/2} |\C|^{1/2}}
\exp \left[-\frac{1}{2} ( \bF(\Sc(\Bs)) - \bd )^T \C^{-1} (
  \bF(\Sc(\Bs)) - \bd )\right],
\label{eq:likelihood}
\end{equation}
where $|\C|$ is the determinant of the covariance $\C$. The estimation
of $\Bs$ is based on the conditional (posterior) density $\bpi(\Bs
|\bd)$.  It is defined by Bayes' rule
\cite{schervish1995theory,fitzpatrick1991bayesian}
\begin{equation}
\bpi(\Bs | \bd) = \frac{\bpi(\Bs,\bd)}{\bpi(\bd)} = \frac{
  \bpi(\bd|\Bs)\ppr(\Bs)}{\bpi (\bd)},
\label{eq:Bayes}
\end{equation}
where $\bpi(\Bs,\bd)$ is the joint probability density of $(\Bs,\bd)$.
The marginal 
\begin{equation}
\label{eq:marginal}
\bpi(\bd) = \int_{\R^g} \bpi(\Bs,\bd) \ppr(\Bs) d \Bs
\end{equation}
is just a normalization that plays no role in the estimation.  The
prior density $\ppr(\Bs)$ may introduce a regularization in the
inverse problem \cite[Chapter 3]{kaipio2005statistical}. The priors
used in our study are summarized in \ref{app:priors}. They all ensure
that the reconstructions $\s(\bx)$ are positive.

We consider maximum a posteriori (MAP) estimates  of the conductivity
\begin{equation}
\sM(\bx) = [\Sc(\BsM)](\bx),
\label{eq:MAP1}
\end{equation}
which maximize the conditional probability density $\pi(\Bs|\bd)$.
The vector $\BsM$ of parameters solves the optimization problem
\begin{equation}
\BsM = \arg \min_{\Bs \in \R^g} [ \bF(\Sc(\Bs)) - \bd ]^T
  \C^{-1} [ \bF(\Sc(\Bs)) - \bd ] - \log(\ppr(\Bs)). 
\label{eq:MAP2}
\end{equation}

The MAP estimates $\BsM$ are random, because they depend on the noise
$\be$ in the measurements. To quantify their uncertainty, we
approximate their variance using a large number $M$ of independent
samples $\BsM^{(m)}$, determined from (\ref{eq:MAP2}) and data
(\ref{eq:data}) with draws $\be^{(m)} \in \mathcal{N}\left({\bf 0},
\C\right)$ of the noise,
\begin{eqnarray}
 \mbox{Var}[\Bs] \approx \frac{1}{M-1} \sum_{m= 1}^M
\left[\BsM^{(m)} - \left< \Bs \right>\right]^2, \quad \left<
\Bs \right> \approx \frac{1}{M} \sum_{m= 1}^M \BsM^{(m)}.
\label{eq:SAMPLES}
\end{eqnarray}
Then, we compare $\mbox{Var}[\Bs]$ to the optimal variance,
which is the right hand side in the Cram\'{e}r-Rao bound
\cite[Corollary 5.23]{schervish1995theory},
\begin{equation}
\mathbb{E}_{\Bs_\star} \{\left[ \BsM - \mathbb{E}_{\Bs_\star}\{\BsM\}
  \right]^2\} \ge {\bf b}^T(\Bs_\star) {\mathcal
  I}^{-1}(\Bs_\star){\bf b}(\Bs_\star).
\label{eq:CRB}
\end{equation}
The notation $\mathbb{E}_{\Bs_\star}$ indicates that the
mean (expectation) depends on the true vector of parameters
$\Bs_\star$. The bias factor ${\bf b}(\Bs)$ is defined by
\begin{equation}
{\bf b}(\Bs)  = \nabla \mathbb{E}_{\Bs}\{\BsM\},
\label{eq:CRBb}
\end{equation}
where $\nabla$ denotes gradient with respect to $\Bs$, and ${\mathcal
  I}(\Bs)$ is the Fischer information matrix \cite[Section
  2.3]{schervish1995theory}. It measures how much information the data
$\bd$ carry about the parameter $\Bs_\star$. The Fisher matrix is in
$\R^{g \times g}$, with entries
\begin{equation}
{\mathcal I}_{i,j}(\Bs) =
\mathbb{E}_{\Bs} \hspace{-0.0in}\left\{ \partial_{s_i} \log
\pi(\bd|\Bs) \, \partial_{s_j} \log \pi(\bd|\Bs) \right\}.
\label{eq:Fisher}
\end{equation}
Since the likelihood $\pi(\bd|\Bs)$ is Gaussian, we obtain under the
natural assumption that the noise covariance $\C$ is independent of
$\Bs$, that
\begin{equation}
{\mathcal I}(\Bs) = D_\Bs \Sc(\Bs)^T D_\s \bF(\Sc(\Bs)) ^T \C^{-1}  D_\s
\bF(\Sc(\Bs)) D_\Bs \Sc(\Bs),
\label{eq:Fisher1}
\end{equation}
where $D_\Bs \Sc(\Bs)$ is the Jacobian of $\Sc(\Bs)$ evaluated at $\Bs$
and $D_\s \bF(\s)$ is the Jacobian of $\bF(\s)$ evaluated at $\s$.
We take as the true $\Bs_\star$ the solution of the optimization
problem (\ref{eq:MAP2}), with noiseless data and upper/lower bound
prior (\ref{eq:ulpr}). The bias ${\bf b}(\Bs_\star)$ is
approximated via Monte Carlo simulations, from a large sample of draws.
The bound (\ref{eq:CRB}) is evaluated and displayed in section
\ref{sect:nonlin}.

\subsection{\textbf{Parametrizations of the conductivity}}
\label{sect:param}
We consider three different parametrizations:
\begin{itemize}
\item {\bf piecewise linear on uniform grid}: The entries in $\Bs$ are
 pointwise values of $\s(\bx)$ on a uniform tensor product grid.  The
 conductivity is piecewise linear on a Delaunay triangulation of the
 grid points.

\item {\bf piecewise linear on optimal grid}: The
 conductivity and the parameters are defined as above, but with grid
 points from the so-called ``optimal grid'' (see
 section~\ref{sect:sensitivity}).

\item {\bf resistor network}: The parameters $\Bs$ are (up to known
 multiplicative constants) the conductors in a network that has the
 same electrical response (DtN map) as the electrical response at the
 electrodes of the unknown conductivity. This parametrization is
 discussed in more detail in section~\ref{sect:nonlinS}.
\end{itemize}

\vspace{0.05in} In the first two parametrizations the conductivity
depends {\em linearly} on $\Bs$. It is the piecewise linear
interpolation of the entries in $\Bs$, between grid nodes
$\bx_1,\ldots,\bx_g$,
\begin{equation}
\s(\bx) = \left[\Sc(\Bs)\right](\bx) = \sum_{i=1}^g s_i
\phi_i(\bx).
\label{eq:Bayes.1}
\end{equation}
Here $\phi_i$ are piecewise linear basis functions on a Delaunay
triangulation of the nodes $\bx_i$, satisfying the usual property
$\phi_{i}(\bx_j) = \delta_{ij}$, with the Kronecker delta notation.

In the resistor network parametrization the conductivity depends {\em
nonlinearly} on the parameters $\Bs$. The dependence is given in
equation (\ref{eq:Bayes4}), in terms of the resistor network reduced
model of (\ref{eq:EIT1}). We emphasize that the resistor network
parametrization cannot be written in the form (\ref{eq:Bayes.1}). One
way of comparing it to the linear case (\ref{eq:Bayes.1}) is to
consider a small perturbation $\delta \Bs \equiv (\delta s_1,
\ldots,\delta s_g)^T$ of some reference $\overline{\Bs}$, and
linearize
\begin{equation}
 \label{eq:Slin}
 \left[\Sc(\overline{\Bs}+\delta \Bs)\right](\bx) =
 \left[\Sc(\overline{\Bs})\right](\bx) + \sum_{i=1}^g \delta s_i
 \phi_i(\bx)+ o(\delta \Bs).
\end{equation}
Here the $\phi_i$ are the ``columns'' of the Jacobian $D_\Bs \Sc
(\overline{\Bs})$ of the parametrization operator $\Sc$ with respect
to $\Bs$, and evaluated at $\overline{\Bs}$. We call these $\phi_i$
{\em sensitivity basis functions}.
\subsection{\textbf{The resistor network discretization and the sensitivity
basis functions}}
\label{sect:nonlinS}
For a given resistor network with $n$ boundary nodes and $g =
n(n-1)/2$ resistors (one per edge), we denote by $\bbF: \real^g_+ \to
\real^g$ the discrete forward map which maps the vector of positive
conductances of the network to a vector of $g$ independent entries of
the Dirichlet-to-Neumann map of the network. The choice and ordering
of the independent entries is identical to that of $\bF$ in
section~\ref{sect:Bayespdf}. We assume that the topology (underlying
graph) of the network is such that $\bbF$ admits a left inverse
$\bbF^{-1}$, i.e. $\bbF^{-1}(\bbF(\bg)) = \bg$ for all $\bg>0$. Such
 topologies are given in section~\ref{sec:solvability}.

In the resistor network parametrization we let
\[
\Bs =(\gamma_1/\gamma_1^{(1)}, \ldots,
\gamma_g/\gamma_g^{(1)})^T \in \R_+^g.
\] 
Here the $\gamma_i$ are the conductances of the network that has the
same electrical response as that measured for $\s$, i.e. $\bF(\s) =
\bbF(\bg)$, for $\bg = (\gamma_1, \ldots, \gamma_g)^T$. Similarly, 
$\bg^{(1)} = (\gamma_1^{(1)}, \ldots, \gamma_g^{(1)})^T$ satisfies $\bF(1) =
\bbF(\bg^{(1)})$. Hence, it is easy to compute $\Bs$ from knowledge of
$\s$ and $\bbF^{-1}$.

The mapping $\Sc$ in the resistor network parametrization is defined
implicitly by
\begin{equation}
 \bF(\Sc(\Bs)) = \bbF(\Bs \bg^{(1)}),
 \label{eq:Bayes3}
\end{equation}
where the product of the vectors $\Bs$ and $\bg^{(1)}$ is
understood componentwise.  There are many functions that satisfy
(\ref{eq:Bayes3}). We define $\Sc(\Bs)$ for the resistor network
parametrization as the limit of the Gauss-Newton sequence
$\{\kappa_j\}_{_{j \ge 0}}$ defined in (\ref{eq:GN}),
\begin{equation}
[\Sc(\Bs)](\bx) = \exp(\kappa(\bx)), \quad \mbox{where} ~~ \kappa = \lim_{j \to
  \infty} \kappa_j.
\label{eq:Bayes4}
\end{equation}
The starting point in the Gauss-Newton iteration is $\kappa_o(\bx) =
\ln(\s^o(\bx))$, where $\s^o(\bx)$ is the piecewise linear interpolation
of the discrete values $\Bs$ on the optimal grid defined in
section~\ref{sect:sensitivity}. In practice, the evaluation of
(\ref{eq:Bayes4}) is computationally efficient because iteration
(\ref{eq:GN}) converges quickly, basically after one step, as shown in
\cite{BorDruGue,GuevaraPhD}.

Recall that for the resistor network parametrization, the mapping $\Sc$ is
non-linear and thus cannot be written as an expansion in basis of
functions, as in (\ref{eq:Bayes.1}). To compare it to parametrizations
of the form (\ref{eq:Bayes.1}), we look at its linearization around
reference parameters $\overline{\Bs}$ to obtain (\ref{eq:Slin}). The
basis functions $\phi_k$ are then the ``columns'' of
$D_\Bs\Sc(\overline{\Bs})$, which can be determined by
differentiating (\ref{eq:Bayes3}) with respect to $\Bs$,
\begin{eqnarray}
 D_\Bs \Sc(\overline{\Bs}) = D_\s\bF^\dagger(\Sc(\overline{\Bs}))
 D_\bg \bbF(\overline{\Bs} \bg^{(1)}) \mbox{diag}(\bg^{(1)}) 
 = (\mbox{diag}(1/\bg^{(1)}) D_\s\bg)^\dagger.
\end{eqnarray}
The sensitivity functions $D_\s\bg(\bx)$ are defined in
section \ref{sect:sensitivity} and are evaluated at $\overline{\s} =
\Sc(\overline{\Bs})$. They give the sensitivity of the resistors
$\bg$ to changes in the conductivity, and are an important ingredient
of the inversion algorithm that is reviewed in section~\ref{sect:Rnets}.

\subsection{\textbf{Outline of the results}}
\label{sect:mainresults}

The linearized problem is studied in section~\ref{sect:lin}. In this
case it it is known \cite[Section 5.1]{schervish1995theory} that the
Cram\'{e}r-Rao bound is attained by the variance of the MAP estimates.
The variance depends of course on the parametrization (\ref{eq:interp})
used in the estimation.  The results say that when we take piecewise
linear interpolations of the parameters $\Bs$ on the equidistant or the
optimal grids, we obtain much larger variances than if we use the
sensitivity basis functions. That is to say, the sensitivity basis
functions lead to better estimates than the piecewise linear ones, even
when we interpolate on the optimal grids. We also show in section
\ref{sect:nonlin} that the nonlinear estimation based on resistor
networks is efficient, in the sense that the sample variance is very
close to the Cram\'{e}r-Rao bound.

\section{EIT with resistor networks}
\label{sect:Rnets}
\setcounter{equation}{0} The resistor networks used in our inversion
algorithms are reduced models of (\ref{eq:EIT1}) that are uniquely
recoverable from discrete measurements of the continuum DtN map, as
described in section \ref{sect:Rnets.R}.  They allow us to estimate
the conductivity $\s$ parametrized on optimal grids, as explained in
section \ref{sect:Rnets.optg}. We give here a brief summary of the
measurement operation, the critical resistor networks, and the induced
reconstruction mapping (\ref{eq:Bayes4}) used in this paper.  We refer
to \cite{BorDruGue,BDM-10,BDMG-10} for more details of the resistor network
based inversion.

\subsection{\textbf{Resistor networks as reduced models for the 
forward and inverse problem}}
\label{sect:Rnets.R}
\begin{figure}[t!]
 \begin{center}
   \includegraphics[width=0.4\textwidth]{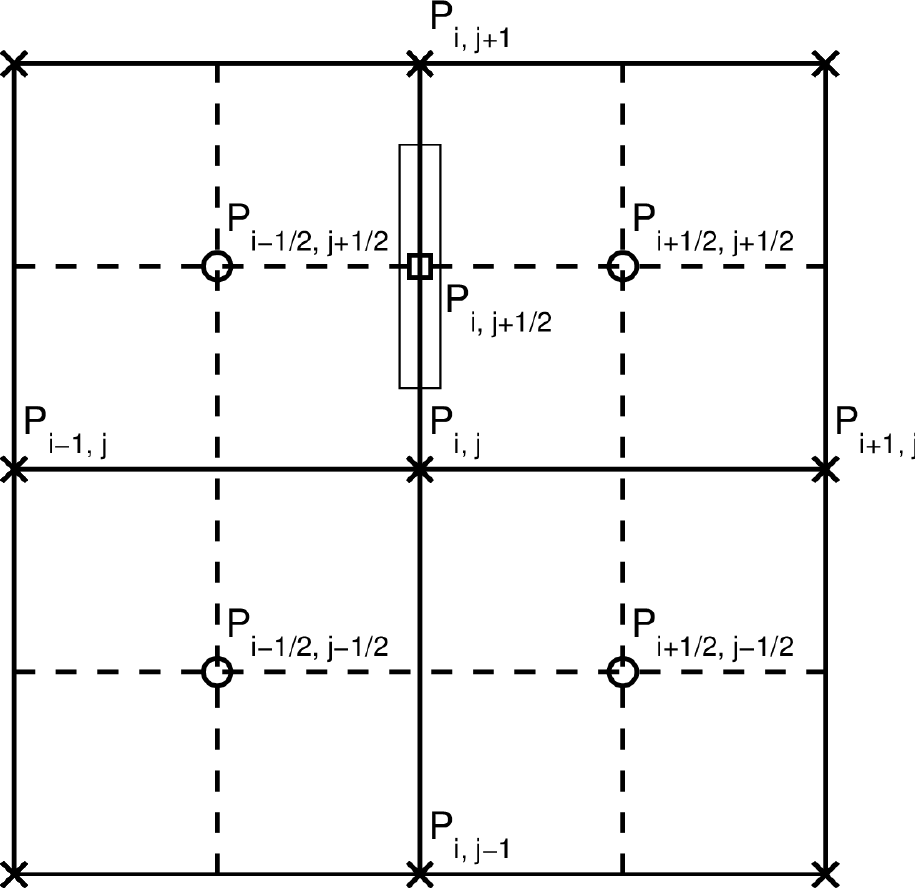}
 \end{center}
 \vspace{-0.1in}\caption[Finite volume discretization]{
         \label{fig:fdgrid}
   \renewcommand{\baselinestretch}{1} \small\normalsize Illustration
   of a staggered grid, with primary (solid) and dual (dashed) grid
   lines.  The primary grid nodes are indicated by $\times$ and the
   dual grid nodes by $\circ$.  We show a resistor as a
   rectangle with a midpoint $\square$ at the intersection of a
   primary and dual line.  }
\end{figure}

Resistor networks arise naturally in finite volume discretizations of
equation (\ref{eq:EIT1}) on staggered grids with interlacing primary
and dual grid lines that may be curvilinear. The potential is
discretized at the primary nodes $P_{i,j}$, the intersections of the
primary grid lines, $u_{i,j} \approx u(P_{i,j}).$ Each node $P_{i,j}
\in \Omega$ is surrounded by a dual cell $C_{i,j}$, as shown in Figure
\ref{fig:fdgrid}. Integrating (\ref{eq:EIT1}) over the cells
$C_{i,j}$, using the divergence theorem, and approximating the
boundary fluxes with finite differences we obtain a system of linear
equations of the form
\begin{equation}
\begin{aligned}
&\gamma_{i+\frac{1}{2},j}(u_{i+1,j}-u_{i,j}) +
 \gamma_{i-\frac{1}{2},j}(u_{i-1,j}-u_{i,j}) \\
+&\gamma_{i,j+\frac{1}{2}}(u_{i,j+1}-u_{i,j}) +
\gamma_{i,j-\frac{1}{2}}(u_{i,j-1}-u_{i,j}) = 0.
\end{aligned}
\label{eq:FVeq}
\end{equation}

Equations (\ref{eq:FVeq}) are Kirchhoff's node law for the interior
nodes of the resistor network with graph $\G = (\Pc,\E)$. Here $\Pc =
\{P_{i,j}\}$ is the set of primary nodes, given by the union of the
disjoint sets $\Pc_{_\B}$ and $\Pc_{_\I}$ of boundary and interior
nodes.  Adjacent primary nodes are connected by edges, the elements of
the set $\E \subset \Pc \times \Pc$.  The network is the pair
$(\G,\bg)$, with $\bg \in \R_+^{|\E|}$ the vector with entries given
by the conductances (inverse of resistances) $\gamma_{\alpha, \beta}
>0$ of the edges, following a preassigned ordering of $\E$.  Here
$(\alpha, \beta) \in \left\{ \left(i,j \pm \frac{1}{2}\right), \;
\left(i \pm \frac{1}{2}, j\right) \right\}$.

We relate the conductances $\gamma_{\alpha, \beta}$ and $\s$ by
\begin{equation}
\gamma_{\alpha, \beta} = \s(P_{\alpha, \beta}) 
\frac{L(\Sigma_{\alpha, \beta})}{ L(E_{\alpha, \beta})},
\label{eq:RAlex}
\end{equation}
where $L$ denotes the arclengths of the primary edges $E_{\alpha,
\beta}$ and dual edges $\Sigma_{\alpha, \beta}$. The points
$P_{\alpha, \beta}$ are located at the intersections of the primary
and dual grid segments $E_{\alpha, \beta}$ and $\Sigma_{\alpha,
\beta}$.  

The \emph{forward problem} for a known network $(\G,\bg)$ amounts to
determining the potential function $\U : \Pc \to \R$, with $u_{i,j} =
\U(P_{i,j})$ satisfying the conservation of currents (\ref{eq:FVeq})
at the interior nodes, and Dirichlet boundary conditions
\begin{equation}
u|_{\Pc_{\B}} = u_{_\B}.
\end{equation}
We denote the number of boundary nodes by $n$. The entries in the vector 
$u_{_\B} \in \mathbb{R}^n$ may be related to the continuum boundary 
potential $V$ as explained below, in section \ref{sect:meas}.  

The \emph{inverse problem} for the network seeks the conductances
$\bg$ from the discrete DtN map $\Lambda_\bg$. The
graph $\G$ is known, and the DtN map is a matrix in $\R^{n \times n}$
that maps the vector $u_{_\B}$ of boundary potentials to the vector
$J_{_\B}$ of boundary current fluxes.  Since we consider the two
dimensional problem, all the graphs $\Gamma$ are {\em circular planar
  graphs} \cite{CurtMooMor,CurtIngMor}, i.e. graphs that can be
embedded in the plane with no crossing edges and with all boundary
nodes $\Pc_{\B}$ lying on a circle.

\subsubsection{\textbf{Discrete measurements of the continuum DtN map.}}
\label{sect:meas}
To connect the discrete inverse problem for the network $(\G,\bg)$ to
continuum EIT, we introduce a \emph{measurement operator} $\M_n$ that
defines a matrix $\M_n \left(\Lambda_\s\right) \in \mathbb{R}^{n
\times n}$ from the continuum DtN map $\Lambda_\s$. The measurement
operator is chosen so that for any suitable conductivity, $\M_n
\left(\Lambda_\s\right)$ is {\em consistent} with the DtN map of a
circular planar resistor network $(\G,\bg)$. The network is a reduced
model for the forward problem, because it satisfies
\begin{equation}
\Lambda_\bg= \M_n \left(\Lambda_\s\right).
\label{eq:meas}
\end{equation}

The continuum forward map $\bF$ of section~\ref{sect:Bayesian} is
defined using this measurement operator as
\begin{equation}
 \bF(\s) = \mbox{vec} (\M_n \left(\Lambda_\s\right)),
\label{eq:meas2}
\end{equation} 
where $\mbox{vec}(A)$ denotes the operation of stacking in a vector in
$\R^g$, $g = n(n-1)/2$, the entries in the strict upper triangular
part of a matrix $A \in \mathbb{R}^{n \times n}$. Because of
reciprocity and conservation of currents, these entries completely
determine the measured DtN map $\M_n\left(\Lambda_\s\right)$.  Hence,
another (equivalent) way of writing the compatibility condition
(\ref{eq:meas}) is
\begin{equation}
 \bbF(\bg) = \bF(\sigma).
\label{eq:meas3}
\end{equation}
where $\bbF(\bg) = \mbox{vec}(\Lambda_\bg)$ is the discrete forward map.

One possible choice of the measurement operator consists of taking
point values of the kernel of $\Lambda_{\sigma}$. Its consistency with
networks is shown in \cite{MorInger,IngerLayer}. Another choice, which
we use in this paper, is to lump fluxes over disjoint segments of $\B$
that model electrode supports. Its consistency with networks is shown
in \cite{BorDruGue, GuevaraPhD}.  Such an operator is defined using
the nonnegative ``electrode'' functions $\chi_1$, $\ldots$, $\chi_n$
in $H^{1/2}(\mathcal{B})$, with disjoint supports, numbered in
circular order on $\mathcal{B}$. We normalize them to integrate to one
on $\B$. The operator $\mathcal{M}_n$ maps $\Lambda_{\sigma}$ to the
symmetric matrix with off-diagonal entries given by
\begin{equation}
(\mathcal{M}_n(\Lambda_\sigma))_{i,j} =
  \langle \chi_i, \Lambda_\sigma \chi_j \rangle, \qquad i \neq j,
\label{eqn:measdtn}
\end{equation}
where $\langle \cdot , \cdot \rangle$ is the duality pairing between
$H^{1/2} (\mathcal{B})$ and $H^{-1/2} (\mathcal{B})$. The diagonal
entries are taken so that the rows (and columns) of $\mathcal{M}_n
(\Lambda_\sigma)$ sum to zero.  Such choice enforces the conservation
of currents. 

\subsubsection{\textbf{Solvability of the inverse problem for resistor 
networks.}}
\label{sec:solvability}
The question of solvability of the inverse problem for circular planar
networks like $(\G,\bg)$ has been settled in
\cite{CurtMooMor,CurtIngMor,IngerLayer,deverdiere1994rep,
deverdiere1996rep}. The answer is that when the graph $\G$ is
\emph{critical}, the discrete forward map $\bbF$ is one-to-one, and
there exists a left inverse $\bbF^{-1}$ so that $\bbF^{-1}(\bbF(\bg))
= \bg$ for all $\bg>0$. A graph is \emph{critical} if it is \emph{well
connected} and if it does not contain any redundant edges.  See
\cite{CurtMooMor,CurtIngMor} for a technical definition of criticality
and well-connectedness. In a critical network, the number $n$ of
boundary nodes and the number $g$ of edges in the graph obey $ g =
{n(n-1)}/{2}.  $ This says that there are as many unknown
conductances in the network as there are degrees of freedom in the DtN
map $\Lambda_\bg \in \R^{n \times n}$.

It remains to define the graph of the network, so that it is critical,
and thus uniquely determined by (\ref{eq:meas}).  Typically, different
graph topologies are better suited for the full and partial data
measurements. Here we use the topologies considered in \cite{BorDruGue,
GuevaraPhD, BDMG-10, MamonovPhD, BDGM-11}, see \ref{app:nets} for
details.

The inverse problem for critical networks $(\G,\bg)$ can be solved
with at least two approaches. We use them both in our study of the
nonlinear inverse problem in section \ref{sect:NumericalRes}.
\begin{enumerate}
 \item {\bf Layer peeling} \cite{CurtMooMor,BDMG-10,BDGM-11}: A direct
 method giving the conductances $\bg$ in a finite number of algebraic
 operations. The advantage of layer peeling is that it is fast and
 explicit. The disadvantage is that it becomes quickly unstable, as
 the size of the network grows. Moreover, noisy data may not be
 consistent with a network, i.e. the consistency relation
 (\ref{eq:meas3}) may not hold if the forward map $\bF$ is noisy.
 \item {\bf Optimization:} Use standard optimization techniques to
find conductances $\bg$ that best fit the (possibly noisy)
measurements in the least squares sense (see section~\ref{sect:rols}
for more details).
\end{enumerate}
\subsection{\textbf{Inversion on optimal grids}}
\label{sect:Rnets.optg}
We denote by $\s_\star$ the true conductivity, to distinguish it from
the estimates that we denote generically by $\s$.  The relations
(\ref{eq:RAlex}) between $\bg$ and $\s_\star$ have been derived in the
discretization of the forward problem. We use them for the
conductances of the network $(\Gamma, \bg)$ recovered from the
measurements $\bF(\s_\star)$, in order to estimate $\s_\star$. This
does not work unless we use a special grid in (\ref{eq:RAlex})
\cite{BorDruGue,BorDruKni}. The idea behind the inversion on optimal
grids is that the geometrical factors $L(\Sigma_{\alpha,
\beta})/L(E_{\alpha, \beta})$ and the distribution of the points
$P_{\alpha, \beta}$ in (\ref{eq:RAlex}) depend weakly on
$\s$. Therefore, we can determine both the geometrical factors and the
grid nodes from the resistor network $(\Gamma,\bg^{(1)})$, with the
same graph $\Gamma$ as before, and $ \bbF(\bg^{(1)}) = \bF(1).  $
These are the measurements of the DtN map for constant conductivity
$\s \equiv 1$, that we can compute, and $ \gamma^{(1)}_{\alpha, \beta}
\approx {L(\Sigma_{\alpha, \beta})}/{ L(E_{\alpha, \beta})}.$ We
obtain the pointwise estimates
\begin{equation}
 \s(P_{\alpha, \beta}) \approx
 \frac{\gamma_{\alpha, \beta}}{\gamma^{(1)}_{\alpha, \beta}},
\label{eq:Step1}
\end{equation} 
that we place in $\Omega$ at points $P_{\alpha, \beta}$ determined
from a sensitivity analysis of the DtN map, as we explain next.

\subsubsection{\textbf{The sensitivity functions and the optimal grids.}}
\label{sect:sensitivity}

The distribution of points $P_{\alpha, \beta}$ in $\Omega$ is
\emph{optimal} in the sense that
\begin{equation}
\bbF(\bg(\s)) = \bF(\s)
\label{eq:SENS1}
\end{equation}
for conductances $\bg = \bg(\s)$ related to the continuum $\s$ as in
(\ref{eq:RAlex}), and for $\s \equiv 1$ (i.e., for $\bg(1) =
\bg^{(1)}$). Each conductance $\gamma_k$ is associated with a point
$P_{\alpha, \beta}$, so we write $k = k(\alpha,\beta)$.  We define the
optimal grid points as the maxima of the sensitivity functions given
below, evaluated at $\s \equiv 1$,
\begin{equation}
P_{\alpha, \beta} = \arg\max_{\bx \in \Omega}\, (D_\sigma
\gamma_{_{k(\alpha, \beta)}})(\bx)\Big|_{\s \equiv 1}.
\label{eqn:sensgriddef}
\end{equation}
These are the points at which the conductances are most sensitive to
changes in the conductivity.

To compute the sensitivity functions, we take derivatives in
(\ref{eq:SENS1}) with respect to $\s$, and obtain 
\begin{equation}
D_\sigma \bg  = \left( D_\bg \bbF(\bg(\sigma))
\right)^{-1} D_\s \bF(\s).
\label{eqn:sensdef}
\end{equation}
The left hand side is a vector-function from $\Omega$ to $\R^{g}$.
Its $k-$th entry is the sensitivity of conductance $\gamma_{k}$ with
respect to changes of $\s$. The matrix $D_\bg \bbF(\bg(\sigma)) \in
\real^{g \times g}$ is invertible \cite{CurtMooMor}. The Jacobian
$D_\s \bF(\s)$ can be written in terms of the Green's function of the
differential operator $u \to \nabla \cdot (\sigma \nabla u)$ with
$u|_{_\B} = 0$, and the ``electrode'' functions $\chi_i$ introduced in
section~\ref{sect:meas}. The calculation is given in detail in
\cite[Section 4]{BDMG-10}.

\subsubsection{\textbf{The estimate of the conductivity on optimal 
grids.}}
\label{sect:invalg}
What we have computed so far allows us to define an initial estimate
$\s^o(\bx)$ of the conductivity, as the linear interpolation of the
values (\ref{eq:Step1}), on the optimal grid defined by
(\ref{eqn:sensgriddef}). Then, we improve the estimate using a
Gauss-Newton iteration that minimizes the objective function
 \begin{equation}
\mathcal{J}(\s) = \left\| \bbQ(\bF(\s)) - \bbQ(\bd) \right\|^2_2
\label{eqn:optimization}
\end{equation}
over search conductivity functions $\sigma(\bx)$.  Here the data vector
is given as in (\ref{eq:data}) for the conductivity $\s_\star$ that we
wish to find. 
The \emph{reconstruction mapping} is 
\begin{equation}
\bbQ(\bd) := \mbox{diag}(1/\bg^{(1)})
\bbF^{-1}(\bd),
\end{equation}
and it involves solving the discrete inverse problem for a resistor
network. The map $\bbQ$ computes the pointwise estimates
(\ref{eq:Step1}) from the data $\bd$. Therefore, $\s^o(\bx)$ is a linear
interpolation of $\bbQ(\bd)$ on the optimal grid.

The objective function $\mathcal{J}(\s)$ is different than the
usual output least squares data misfit $\left\| \bF(\s) - \bd
\right\|^2_2$. We use $\bbQ$ in (\ref{eqn:optimization}) as a {\em
nonlinear preconditioner} of the forward map $\bF$, as explained in
detail in \cite{BorDruGue, GuevaraPhD}. It is because of this
preconditioning, and the good initial guess $\s^o(\bx)$, that we can
obtain close estimates of $\s$ by minimizing $\mathcal{J}(\s)$
in \cite{BorDruGue, GuevaraPhD} and in this paper. The estimates are
computed with a Gauss-Newton iteration that basically converges in one
step \cite{BorDruGue, GuevaraPhD}.

We enforce the positivity of $\s$ by the change of variables 
$\kappa = \ln(\s)$, so that we work with the map 
\begin{equation}
\bG(\kappa) = \ln \left[\bbQ ( \bF(\exp(\kappa)) )\right].
\label{eq:mapG}
\end{equation} 
The Gauss-Newton iteration that we use in the parametrization
(\ref{eq:Bayes4}) is 
\begin{equation}
\kappa_j = \kappa_{j-1} + \left(D_\kappa \bG (\kappa_{j-1})
\right)^\dagger \left[ \ln [\bbQ(\bd)] - \bG(\kappa_{j-1}) \right],
\quad j = 1, 2, \ldots,
\label{eq:GN}
\end{equation}
with initial guess $\kappa_o = \ln(\s^o)$. The index $\dagger$ denotes
the Moore-Penrose pseudo-inverse, and the iteration amounts to finding
the update $\kappa_j - \kappa_{j-1}$ as the orthogonal projection of
the residual onto the span of the sensitivities, the column space of
the transpose of $D_\kappa \bG(\kappa_{j-1})$. These sensitivities are
easily related to those computed in section \ref{sect:sensitivity},
using the chain rule to deal with the change of variables $\kappa =
\ln (\s)$.

\section{Numerical experiments setup}
\label{sect:setup}
We explain in section \ref{sect:noise} how we simulate the noisy
measurements. The noise may be too high for the layer peeling method
to work. The optimization method presented in section \ref{sect:rols}
is more robust to noise and, as a bonus, it allows us to solve
efficiently the optimization problem for the MAP estimate
(\ref{eq:MAP2}) with the resistor network discretization (see
Remark~\ref{rem:exact}).  For reference, we include noiseless
reconstructions in section \ref{sect:noiseless}.

\subsection{\textbf{Data and noise models}}
\label{sect:noise}
We solve equation (\ref{eq:EIT1}), with a second order finite volume
method on a very fine, uniform, tensor product grid, with $N \gg n$
nodes on the boundary. This approximates $\M_N(\Lambda_{\s_\star})$,
where $\M_N$ is the measurement operator of section~\ref{sect:meas}.
The noise $\be$ in (\ref{eq:data}) is given by 
\begin{equation}
\be = \mbox{vec} \left( \M_n(\bEc)\right), \qquad \bEc \in
\mathbb{R}^{N \times N}.
\label{eq:N1}
\end{equation}

We use two noise models, defined in terms of the noise level $\ell$
and a symmetric $N \times N$ matrix $\boeta$, with Gaussian,
identically distributed entries with mean zero and variance one. The
entries in $\boeta$ on and above the diagonal are uncorrelated.  The
first model scales the noise by the entries of the DtN map for
constant conductivity $\s \equiv 1$,
\begin{equation}
\label{eq:multN}
\bEc = \ell \, \M_N(\Lambda_1) \cdot \boeta,
\end{equation} 
where symbol $\cdot$ stands for componentwise multiplication. The
scaling makes the noise easier to deal with, and the model is somewhat
similar to {\em multiplicative noise}.  The second noise model is
\begin{equation}
\label{eq:addN}
\bEc = \ell \frac{\boeta}{\|\boeta\|},
\end{equation}
where $\|~\|$ is a matrix norm that approximates the continuum
$H^{1/2}(\B) \to H^{-1/2}(\B)$ operator norm. It is defined in
\ref{app:norm}.

\subsection{\textbf{The Gauss-Newton iteration for determining 
the resistor networks}}
\label{sect:rols}
The direct, layer peeling algorithms described in
\cite{CurtMooMor,BDMG-10,BDGM-11} are fast, but highly unstable and
can be used only for very small noise. It is not known how to
regularize layer peeling algorithms. To deal with the instability, we
can only reduce the size of the network, as was done in
\cite{BorDruGue}.  However, simply reducing the network size is not
sufficient for the larger noise levels considered in the simulations.
We use instead the more robust Gauss-Newton method described below,
which allows regularization. The Gauss-Newton method is more expensive
than layer peeling (about 20 times more expensive for $n=7$), but the
computational cost is reasonable because the dimension $g$ of the
vector of unknown conductances is small for noisy data, and the
Jacobian is relatively inexpensive to compute.

The Gauss-Newton method determines the log-conductances $\bkappa =
\ln(\bg)$ in the network $(\Gamma,\bg)$ (with topology $\Gamma$ fixed)
by minimizing the objective functional
\begin{equation}
\mathcal{O}(\bkappa) = (\bbF(\exp(\bkappa)) - {\bf d})^T
\C^{-1} (\bbF(\exp(\bkappa)) - {\bf d}) + \alpha || \bkappa -
\bkappa_\reff ||^2,
\label{eq:rols}
\end{equation}
over $\bkappa \in \mathbb{R}^g$. Thus, the positivity of the
conductances $\bg = \exp(\bkappa)$ is satisfied automatically. The
first term in (\ref{eq:rols}) measures the misfit between the measured
data ${\bf d}$ modeled by (\ref{eq:data}), and the data produced by a
network with log-conductances $\bkappa$.  Here $\C$ is the covariance
matrix of the measurements. The second term in (\ref{eq:rols}) is a 
Tikhonov type regularization penalizing the distance from some 
reference log-conductances $\bkappa_\reff$. The parameter $\alpha>0$ 
determines the strength of the penalty term.

In our numerical experiments, the Gauss-Newton approximation of the
Hessian of $\mathcal{O}(\bkappa)$ is further regularized by
adding $10^{-4}/\C_{1,1}$ to its diagonal.  The iterations are stopped
either when the norm of the gradient of $\mathcal{O}(\bkappa)$
is $10^{-4}$ smaller than that at the initial iterate, or when the
maximum number of iterations (300) is reached. The initial iterate is
$\bkappa = \ln(\bg^{(1)})$, and we take as reference log-conductances
$\bkappa_\reff = \ln(\bg^{(1)})$.

\begin{remark} 
\label{rem:exact}
Recall from (\ref{eq:Bayes3}) that the network is the reduced model
that matches the data. Therefore:
\begin{enumerate}
 \item When $\alpha>0$, finding log-conductances that minimize 
 $\mathcal{O}(\bkappa)$ is equivalent to finding the MAP estimate
 (\ref{eq:MAP2}), with the resistor network based parametrization, and
 the prior (\ref{eq:gppr}). The regularization parameter $\alpha$ is
 the same in both (\ref{eq:MAP2}) and (\ref{eq:rols}). Moreover, when
 $\bkappa_\reff = \ln(\bg^{(1)})$ in (\ref{eq:rols}), the reference
 parameters appearing in (\ref{eq:gppr}) are $\Bs_\reff = \bone$
 (vector of all ones with length $g$).
 \item When $\alpha=0$, minimizing $\mathcal{O}(\bkappa)$ is equivalent to
   finding the MAP estimate (\ref{eq:MAP2}) with the
   resistor network based parametrization and the
   upper/lower bound (\ref{eq:ulpr}) (for a large enough upper bound
   $\s_\mathrm{max}$).
\end{enumerate}
\end{remark}

\subsection{\textbf{Noiseless reconstructions}}
\label{sect:noiseless}
\begin{figure}[tb]
 \begin{center}
   \includegraphics[width=0.5\textwidth]
   {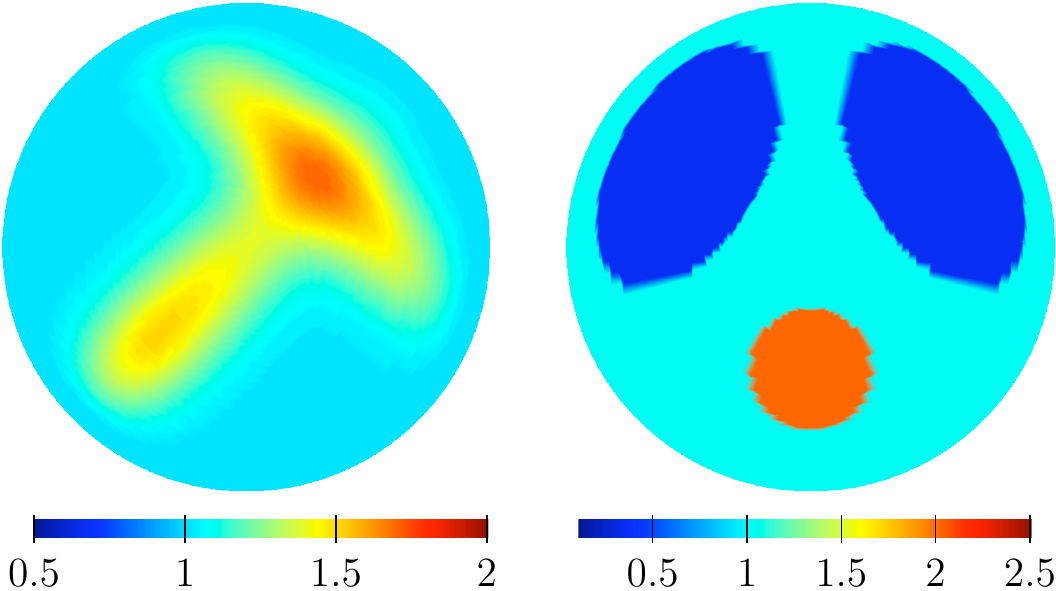}
\end{center}
\vspace{-0.1in} \caption[]{\label{fig:Phantoms}
   \renewcommand{\baselinestretch}{1} \small\normalsize
 Conductivities used in the numerical experiments. 
Left: smooth conductivity. Right: piecewise constant chest phantom.
}  
\end{figure}

\begin{figure}[tb]
 \begin{center}
 \includegraphics[width=0.8\textwidth]{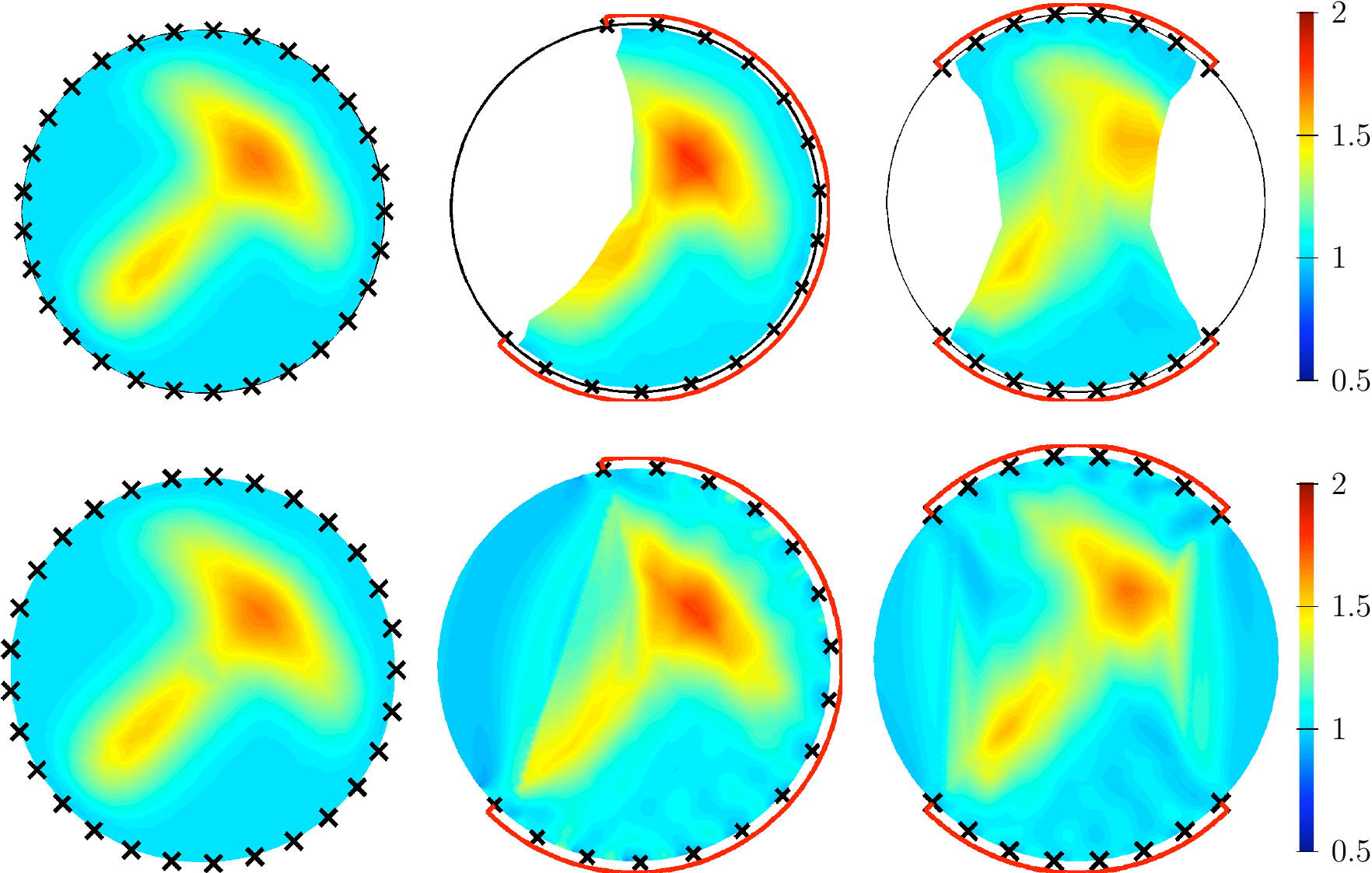}
 \end{center}
\vspace{-0.1in} \caption[]{\label{fig:smoothNoiseless}
   \renewcommand{\baselinestretch}{1} \small\normalsize
   Reconstructions of the smooth conductivity with noiseless data.
   }
\end{figure}

\begin{figure}[h]
\begin{center}
 \vspace{0.2in}\includegraphics[width=0.8\textwidth]{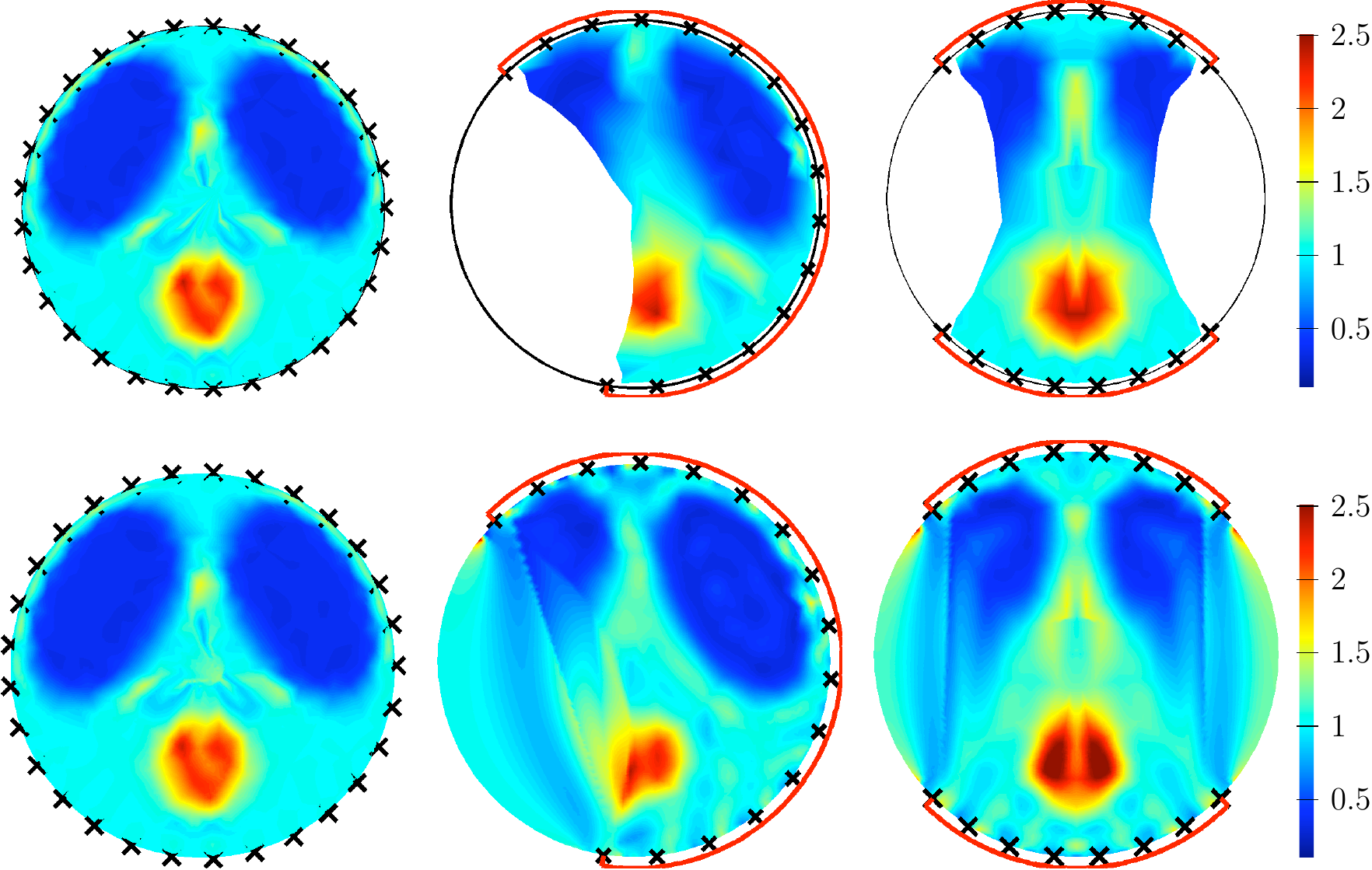}
\end{center}
 \vspace{-0.1in}\caption[]{\label{fig:CHESTNoiseless}
   \renewcommand{\baselinestretch}{1} \small\normalsize
   Reconstructions of the piecewise constant conductivity with
   noiseless data.  Here and in Figure \ref{fig:smoothNoiseless}: Left
   column: full boundary measurements. Middle column: one-sided
   partial measurements. Right column: two-sided partial boundary
   measurements. The top row shows the initial guess $\s^o(\bx)$. The
   bottom row shows the result of one step of the Gauss-Newton
   iteration. The color scale is the same used for the true
   conductivity in Figure \ref{fig:Phantoms}.}
\end{figure}

We show in Figures \ref{fig:smoothNoiseless} and
\ref{fig:CHESTNoiseless} the reconstructions of the true smooth and
piecewise constant conductivities displayed in Figure
\ref{fig:Phantoms}.  The reconstructions are obtained in the full and
partial boundary setups, and for noiseless data. Two distinct cases of
partial boundary measurements are considered. In the \emph{one-sided}
case the accessible boundary $\B_A$ consists of a single connected
segment of $\B$. In the \emph{two-sided} case the accessible boundary
$\B_A$ consists of two disjoint segments of $\B$.

We display in the top row of Figures \ref{fig:smoothNoiseless} and
\ref{fig:CHESTNoiseless} the initial guess $\s^o(\bx)$ of the
Gauss-Newton iteration (\ref{eq:GN}). 
It is the piecewise linear
interpolation of the values $\s^o(P_{\alpha,\beta}) =
\s_{k(\alpha,\beta)}$, where $\s_k$ are obtained from (\ref{eq:Step1}),
and the optimal grid nodes $P_{\alpha,\beta}$ are defined by
(\ref{eqn:sensgriddef}). The function $\s^o(\bx)$ is linear on the
triangles obtained by a Delaunay triangulation of the points
$P_{\alpha,\beta}$.  
In the partial boundary measurements case we display $\s^o(\bx)$ in
the subdomain delimited by the accessible boundary and the segmented
arc connecting the innermost grid points. We set $\s^o(\bx)$ to the
constant value one in the remainder of the domain.  The plots in the
bottom row in Figures \ref{fig:smoothNoiseless} and
\ref{fig:CHESTNoiseless} display the result of one step of the
Gauss-Newton iteration described in section \ref{sect:invalg}.

The network topologies are defined in \ref{app:nets}. The
reconstructions from full boundary data are in the left column in
Figures \ref{fig:smoothNoiseless} and \ref{fig:CHESTNoiseless}. They
are obtained with a circular network $C\left(\frac{n-1}{2},n\right)$,
with $n = 29$.  The reconstructions with the one-sided partial
boundary measurements are in the middle column, and they are obtained
with a pyramidal resistor network $\Gamma_n$, for $n=16$. The
reconstructions with the two-sided boundary measurements are in the
right column. They are obtained with a resistor network $T_n$, for $n
= 16$.

\section{Effect of parametrization on the reconstruction error}
\label{sect:NumericalRes}

We study numerically the effect of noise on the inversion with (a)
resistor networks, and (b) the conductivity parametrized with
piecewise linear basis functions. We consider in section
\ref{sect:lin} the linearized problem about the constant conductivity
$\os \equiv 1$, and in section \ref{sect:nonlin} the nonlinear
problem.

\subsection{\textbf{The linearized problem}}
\label{sect:lin}

\begin{figure}[htpb]
\begin{center}
\includegraphics[width=0.8\textwidth]{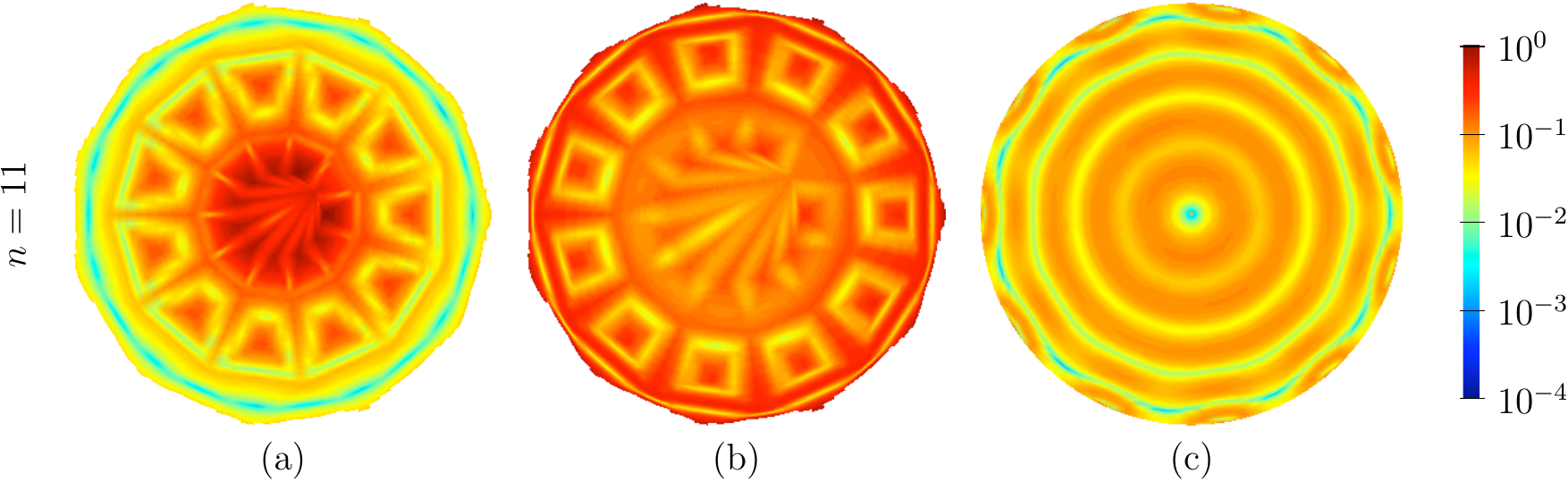}
\end{center}
\vspace{-0.1in}
\caption[]{\label{fig:LIN1} \renewcommand{\baselinestretch}{1}
  \small\normalsize Standard deviation of reconstructions for the
  linearized problem.  (a): Piecewise linear basis functions on
  uniform grid. (b): Piecewise linear basis functions on optimal
  grid. (c): Resistor network parametrization.  The color scale is
  base-10 logarithmic and the noise level is $\ell = 0.01\%$ (additive
  noise model (\ref{eq:addN})). The circular resistor network has $n =
  11$ boundary points and the noise $\bEc$ is given at $N = 100$
  equidistant points on the boundary.}
\end{figure}

The results in this section are for linearization at the constant
conductivity $\os(\bx) \equiv 1$, with additive noise modeled as in
(\ref{eq:addN}). All the three parametrizations described in section
\ref{sect:param} represent exactly $\os \equiv 1$, with a parameter
vector $\Bs$ of all ones, i.e.  $\Sc(\bone) = 1$.  Hence, the
linearization of the forward map around $\overline{\Bs} = 1$ can be
written as
\begin{equation}
 \bF (\Sc(\bone + \delta \Bs)) =  \bF(1) + D_\s\bF(1) D_\Bs
 \Sc(\bone) \delta \Bs + o(\delta\Bs),
 \label{eq:linbF}
\end{equation}
for a small perturbation $\delta\Bs$ of the parameters. If we
discretize the conductivity on a fine grid with $N_\s$ points, the
Jacobian $D_\s \bF(1)$ is a $g \times N_\s$ matrix, and the Jacobian
$D_\Bs \Sc(\bone)$ is an $N_\s \times g$ matrix with columns given by
the basis functions $\phi_k$ in (\ref{eq:Slin}).

The estimate $\sM$ is calculated by solving the optimization problem
(\ref{eq:MAP2}), with the linearization (\ref{eq:linbF}) of the
forward map, and the upper/lower bound prior 
(\ref{eq:ulpr}).  This optimization is a quadratic
programming problem that we solve using the software QPC
\cite{willsQPCweb}.  The mean and variance of $\sM$ are estimated with
Monte Carlo simulations
\begin{eqnarray}
 \mbox{Var}[\s(\bx)] \approx \frac{1}{M-1} \sum_{m= 1}^M
\left[\sM^{(m)}(\bx) - \left< \s(\bx) \right>\right]^2, \quad \left<
\s(\bx) \right> \approx \frac{1}{M} \sum_{m= 1}^M \sM^{(m)}(\bx),
\label{eq:CSAMPLES}
\end{eqnarray}
using $M = 1000$ samples.  We do not show the mean
$\left<\s(\bx)\right>$ because it is basically $\os(\bx)$ for all the
cases that we present below. The standard deviation
$\left\{\mbox{Var}[ \s(\bx)]\right\}^{1/2}$ is shown in
Figure~\ref{fig:LIN1} for the case of full boundary measurements, and
the three parametrizations described in section \ref{sect:param}.  The
noise level is $\ell = 0.01\%$. We choose it so small to minimize the
action of the positivity constraints imposed by the prior. Larger
noise levels are considered later in the paper.  

With the resistor network parametrization, the standard deviation is
smaller and does not increase toward the center of the domain.  Also
there are no active positivity constraints.  The random fluctuations
of $\sM$ in Figure~\ref{fig:LIN1}~(b) lie mostly within three standard
deviations, and are much smaller than the background conductivity $\os
\equiv 1$.

The piecewise linear parametrization on the equidistant grid gives a
large, order one standard deviation in the center of the domain.  The
positivity constraints are active in $81.6\%$ of
realizations. Surprisingly, the piecewise linear parametrization on
the optimal grid is worse. Its standard deviation is large, of order
one in most of the domain, and the positivity constraints are active
in $61.1\%$ realizations.  This shows that it is not enough to
distribute the parameters on the optimal grid.

\begin{figure}[tb]
\begin{center}
\begin{tabular}{c@{}c}
\includegraphics[height=0.22\textheight]{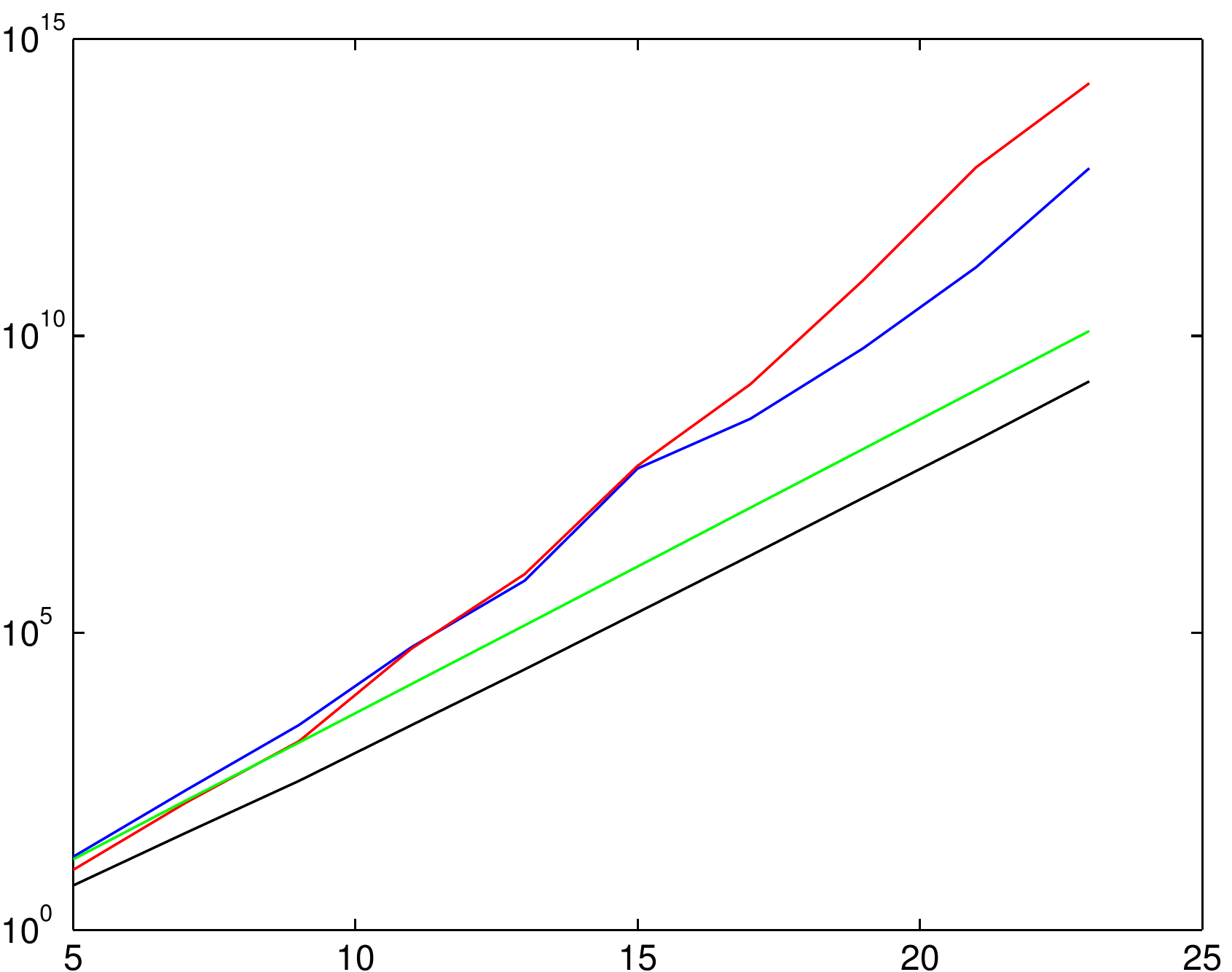}
\end{tabular}
\end{center}
\vspace{-0.1in} \caption[]{\label{fig:LIN2} 
  \renewcommand{\baselinestretch}{1} \small\normalsize Comparison of the
  condition number of the matrix $D_\s \bF(1) D_\Bs \Sc (\bone)$.  The
  abscissa is the number $n$ of boundary nodes and the ordinate is in
  logarithmic scale. The green line is for the fine grid forward map
  ($\Sc =$ identity). The red line is for the piecewise linear functions
  on the equidistant grid. The blue line is for the
  piecewise linear functions on the optimal grid. The black line is for
  the resistor network parametrization.}
\end{figure}

The same conclusion can be reached from Figure \ref{fig:LIN2}, where we
display the condition number of the matrix $D_\s {\bf F}(1) D_\Bs
\Sc(\bone)$, as a function of the number $n$ of boundary points. The
condition number increases exponentially with $n$, as expected from the
exponential ill-posedness of the problem. However, the rate of increase
is smaller for the resistor network parametrization.

\begin{figure}[tb]
 \begin{center}
 \includegraphics[width=0.48\textwidth]{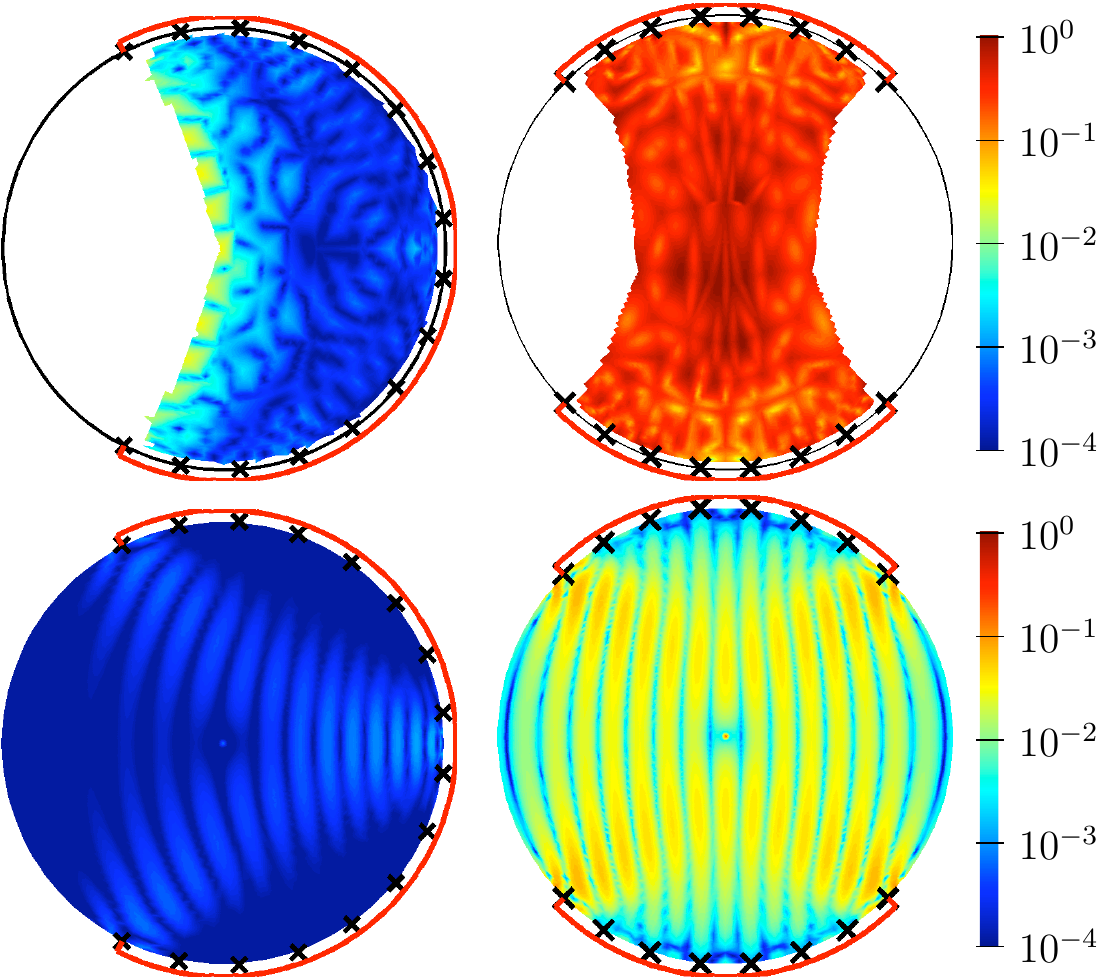}
 \end{center}
 \vspace{-0.1in} \caption[]{\label{fig:prmdrecvar} \renewcommand{\baselinestretch}{1}
   \small\normalsize The standard deviation (base-10 log scale) for
   one-sided (left) and two-sided (right) boundary measurements. The
   resistor networks have $n = 16$ boundary points and the noise
   $\bEc$ with level $\ell = 10^{-8}\%$ is given at $N = 83$ (left)
   and $N = 104$ (right) equidistant points on the accessible
   boundary.  Top: piecewise linear parametrization on the optimal
   grid. Bottom: resistor network parametrization. The accessible
   boundary is  in solid red.  }
\end{figure}

 The standard deviation $\left\{\mbox{Var}[ \s(\bx)]\right\}^{1/2}$ is
 shown in Figure \ref{fig:prmdrecvar} for the case of one and
 two-sided boundary measurements. We use a much smaller noise level
 ($\ell = 10^{-8}\%$, additive model (\ref{eq:addN})) in the partial
 measurements case than in the full data case, because we compute the
 standard deviation for bigger networks ($n=16$, same network size as
 in the reconstructions of Figures \ref{fig:smoothNoiseless} and
 \ref{fig:CHESTNoiseless}). Since the condition number of the
 linearized problem grows exponentially as we increase $n$, only very
 small levels of noise can be used for $n = 16$.  We present the
 results for the piecewise linear parametrization on the optimal grid
 (top row) and the resistor network parametrization (bottom row).
 
 We reach the same conclusion as before. The resistor network
 parametrization gives a smaller standard deviation, that does not
 increase toward the inaccessible region. The piecewise linear
 parametrization on the optimal grid gives a large standard deviation
 near the inaccessible region in the one-sided case, and in the whole
 domain in the two-sided case. The positivity constraints are active
 in most realizations for the piecewise linear parametrization.  They
 are not active for the resistor network parametrization.

\subsection{\textbf{The nonlinear problem}}
\label{sect:nonlin}

We study the statistics (mean and standard deviation) of the MAP
estimates of the parameters $\BsM$ and the conductivities $\sM =
\Sc(\BsM)$, which come from minimizing the functional (\ref{eq:MAP2}).
The study can be done for both the full and partial boundary
measurement setup, but we present here only the full measurements
case. We consider first, in section~\ref{sect:direct}, very small
noise so that we can use the fast layer peeling inversion algorithm to
find the minimizer of (\ref{eq:MAP2}), with resistor network
parametrization and upper/lower bound prior (\ref{eq:ulpr}).  The
speed of the algorithm allows us to compute the Cram\'er-Rao lower
bound in a reasonable amount of time.  We do not calculate the bound
for larger noise, where we use regularized Gauss-Newton to determine
the resistors, because of the computational cost. However, we do show
in section \ref{sect:gn} the bias and relative standard deviation of
the reconstructions, and we also compare in section \ref{sect:lcurve}
the results to those with a piecewise linear discretization and
Gaussian prior on the conductivity (\ref{eq:gcpr}).

\subsubsection{\textbf{Statistics of resistor network inversion using
layer peeling}}
\label{sect:direct}

The results in this section are for the MAP estimates $\BsM$, solving
the optimization problem (\ref{eq:MAP2}), with the resistor network
parametrization, and upper/lower bound prior (\ref{eq:ulpr}). We
present the mean $\left< \Bs \right>$,  $\mbox{Bias}[\Bs] =
\Bs_\star - \left< \Bs \right>$, and the standard deviation
$(\mbox{Var}[\Bs])^{1/2}$ of the estimates. We consider a very small
noise level $\ell=0.1\%$ (noise model (\ref{eq:multN})), so that we
are able to minimize (\ref{eq:rols}) with no regularization
($\alpha=0$) directly, using the layer peeling algorithm.

Note that in the Cram\'er-Rao bound, the Fischer matrix
(\ref{eq:Fisher}) can be calculated analytically, but the bias factor
(\ref{eq:CRBb}) is estimated with Monte Carlo simulations. This is the
expensive part of the computation, because we need a large number of
samples to estimate the mean. In addition, each component of the vector
$\Bs$ is perturbed to approximate the partial derivatives in
(\ref{eq:CRBb}) via finite differences.  We use $M = 1000$ samples, and
the bias is relative to $\Bs_\star$, the solution of the optimization
problem (\ref{eq:MAP2}) with noiseless data. The partial derivatives in
(\ref{eq:CRBb}) are approximated with finite differences with a step
size of $0.01$.

\begin{figure}[tb]
\begin{center}
\includegraphics[width=0.4\textwidth]{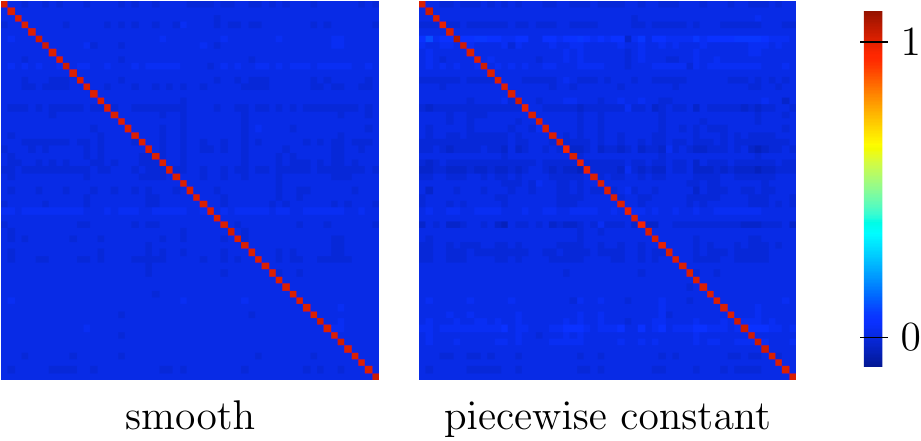}
\end{center}
\vspace{-0.1in} \caption[]{\label{fig:bias} 
\renewcommand{\baselinestretch}{1} \small\normalsize
Estimated bias factor (\ref{eq:CRBb}) for conductivities in Figure
\ref{fig:Phantoms} is close to the identity.}
\end{figure}
\begin{figure}[tb]
\begin{center}
\includegraphics[width=0.8\textwidth]{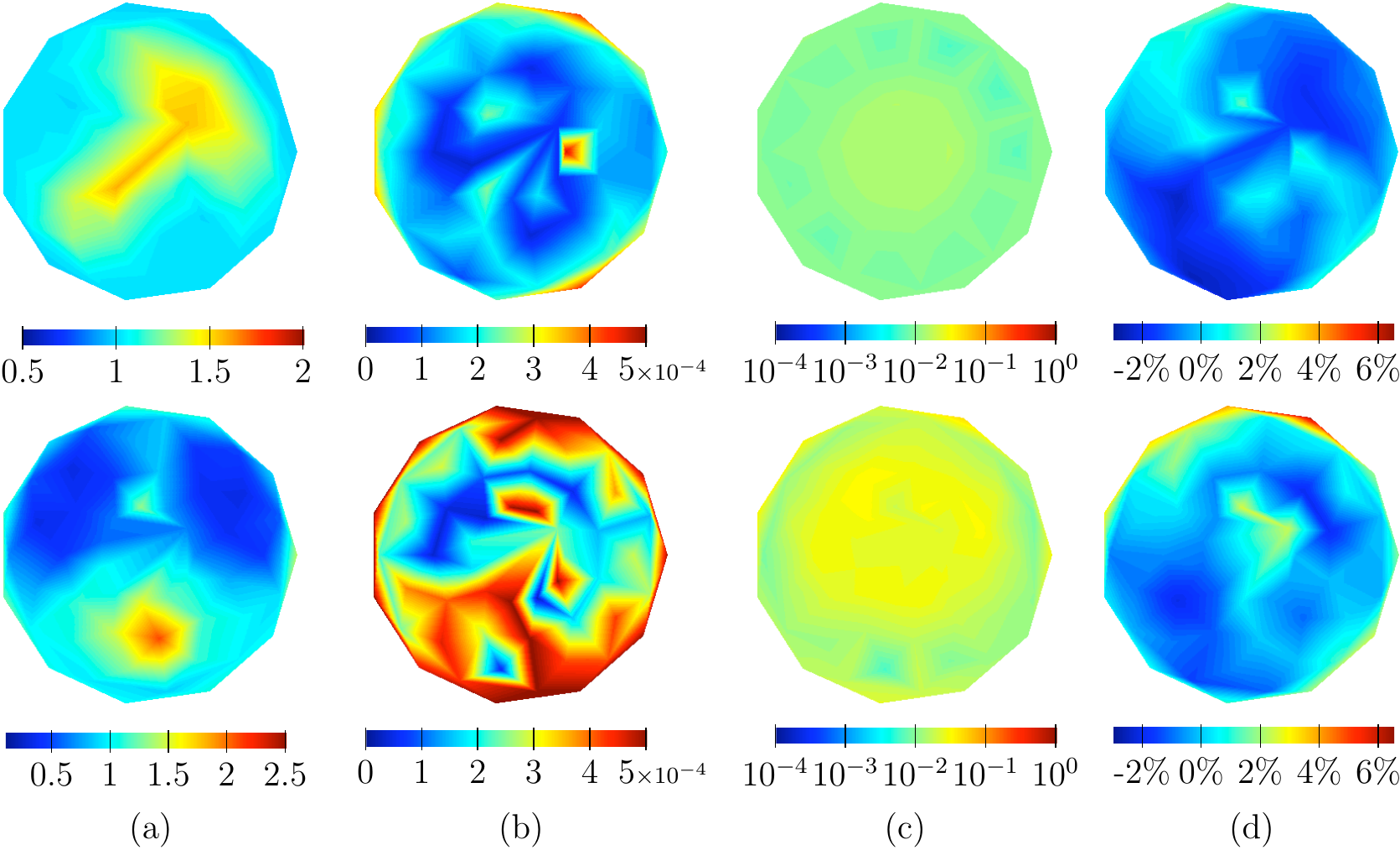}
\end{center}
\vspace{-0.1in} \caption[]{\label{fig:CRB} \renewcommand{\baselinestretch}{1}
\small\normalsize Results with the direct resistor finding algorithm:
(a) mean, (b) absolute bias, (c) standard deviation relative to the
mean, (d) relative difference between variance and Cram\'er-Rao bound
relative to Cram\'er-Rao bound (in percentage).  The noise model is
(\ref{eq:multN}) and the level is $\ell = 0.1\%$. The circular
resistor network has $n = 11$ boundary points and the noise matrix
$\bEc$ is given at $N = 100$ equidistant points on $\B$. All
parameters are linearly interpolated on the optimal grid.  }
\end{figure}


The bias factor is shown in Figure \ref{fig:bias}, and it is close to
the identity matrix. That is to say, the estimates are
unbiased. Figure \ref{fig:CRB} shows (a) the mean $\left< \Bs
\right>$, (b) $\mbox{Bias}[\Bs]$ and (c) the relative standard
deviation $(\mbox{Var}[\Bs])^{1/2}/\left< \Bs \right>$, where the
division is understood componentwise. The last column (d) shows the
difference in percentage between $\mbox{Var}[\Bs]$ and the
Cram\'er-Rao bound in (\ref{eq:CRB}), normalized pointwise by the
Cram\'er-Rao bound. We evaluate the Cram\'er-Rao bound by setting the
bias factor (\ref{eq:CRBb}) to the identity, which is a good
approximation (recall Figure~\ref{fig:bias}). Note that the difference
between the variance and the Cram\'er-Rao bound is very small,
indicating that the estimation is efficient. The result in column (d)
should be non-negative.  We have some negative numbers, probably due
to insufficient sampling in Monte Carlo, but they are so small in
absolute value that we can treat them as essentially zero.

\subsubsection{\textbf{Statistics of conductivity estimates using
optimization}}
\label{sect:gn}

\begin{figure}[tb]
\begin{center}
\includegraphics[width=0.58\textwidth]{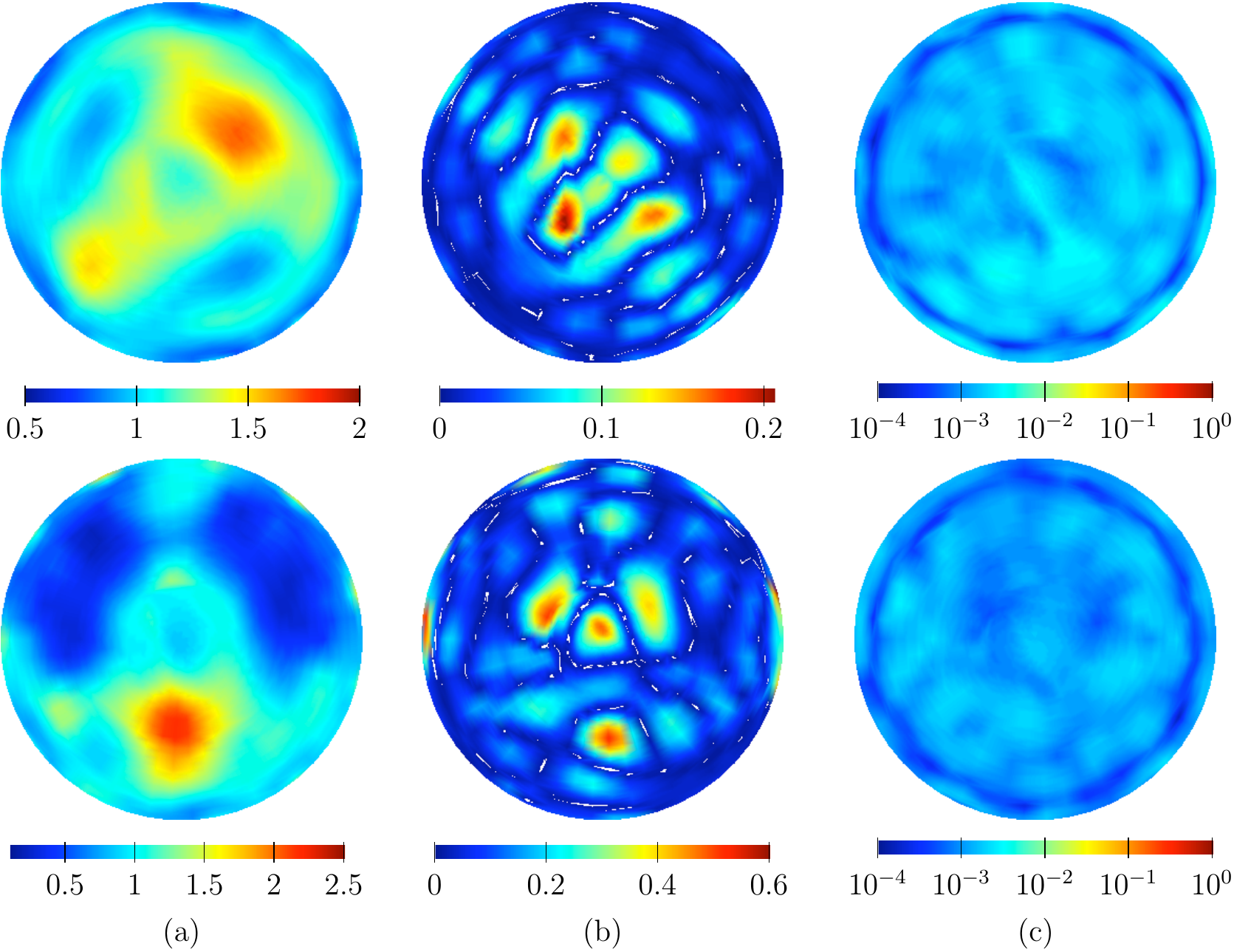}
\end{center}
\vspace{-0.1in} \caption[]{\label{fig:rols11} 
\renewcommand{\baselinestretch}{1} \small\normalsize
Results for $\ell = 0.01\%$ additive noise
(\ref{eq:addN}) and a regularization parameter $\alpha=10^{-6}$ in
(\ref{eq:rols}). The circular resistor network has $n = 11$ boundary
points and $\bEc$ is given at $N = 100$ equidistant points on $\B$.
Top: the smooth conductivity shown on the left in Figure
\ref{fig:Phantoms}. Bottom: the piecewise constant conductivity shown
on the right in Figure \ref{fig:Phantoms}. (a) mean, (b) bias, (c)
relative standard deviation. 
}
\end{figure}

\begin{figure}[tb]
\begin{center}
\includegraphics[width=0.55\textwidth]{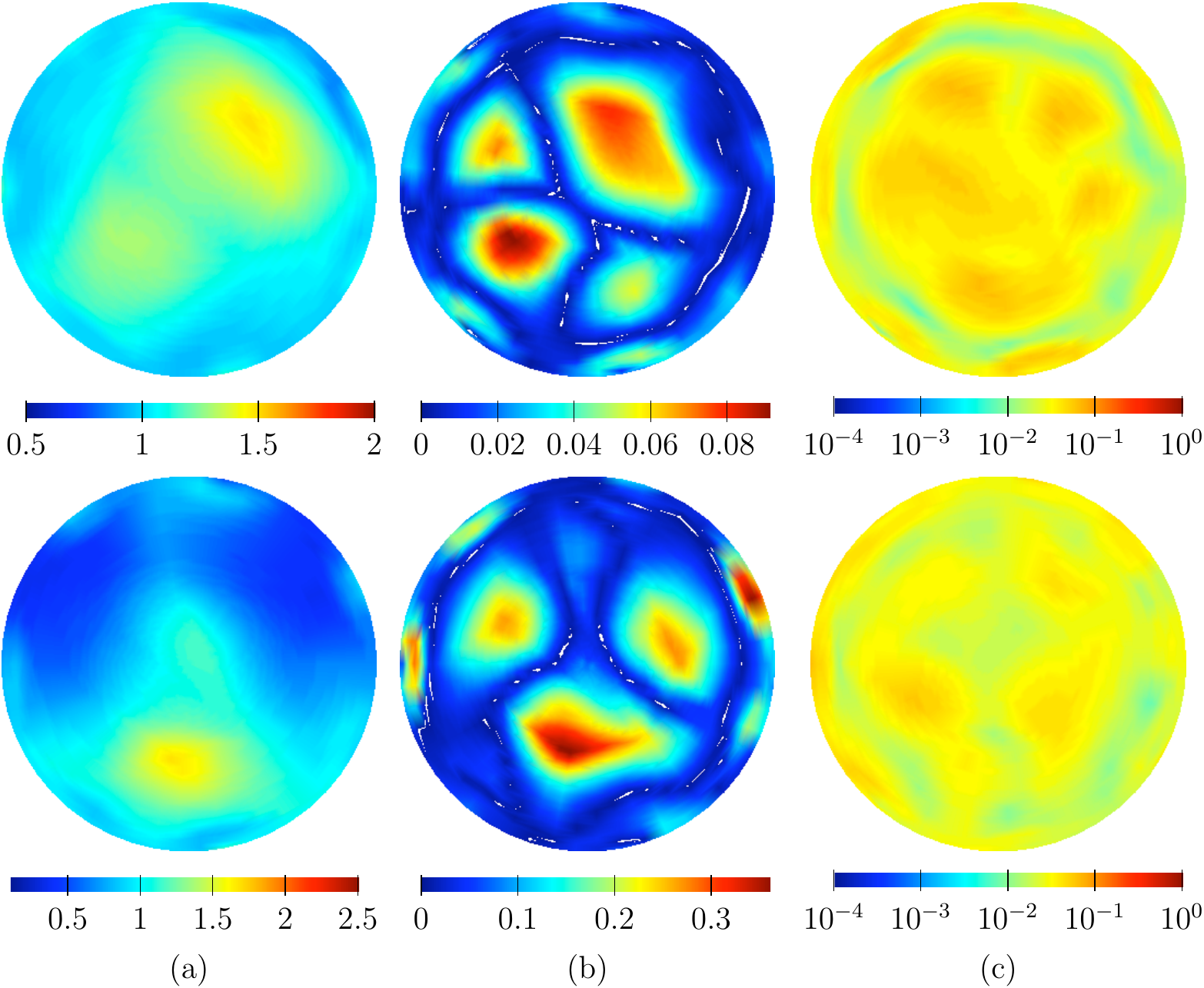}
\end{center}
\vspace{-0.1in} \caption[]{\label{fig:rols7} 
\renewcommand{\baselinestretch}{1} \small\normalsize
Results for $\ell = 1\%$ additive noise
(\ref{eq:addN}) and a regularization parameter $\alpha=10^{-3}$ in
(\ref{eq:rols}). The circular resistor network has $n = 7$ boundary
points and $\bEc$ is given at $N = 100$ equidistant points on $\B$.
Top: the smooth conductivity shown on the left in Figure
\ref{fig:Phantoms}. Bottom: the piecewise constant conductivity shown
on the right in Figure \ref{fig:Phantoms}.(a) mean, (b) bias, (c)
relative standard deviation. 
}
\end{figure}

\begin{figure}
\begin{center}
\includegraphics[width=0.8\textwidth]{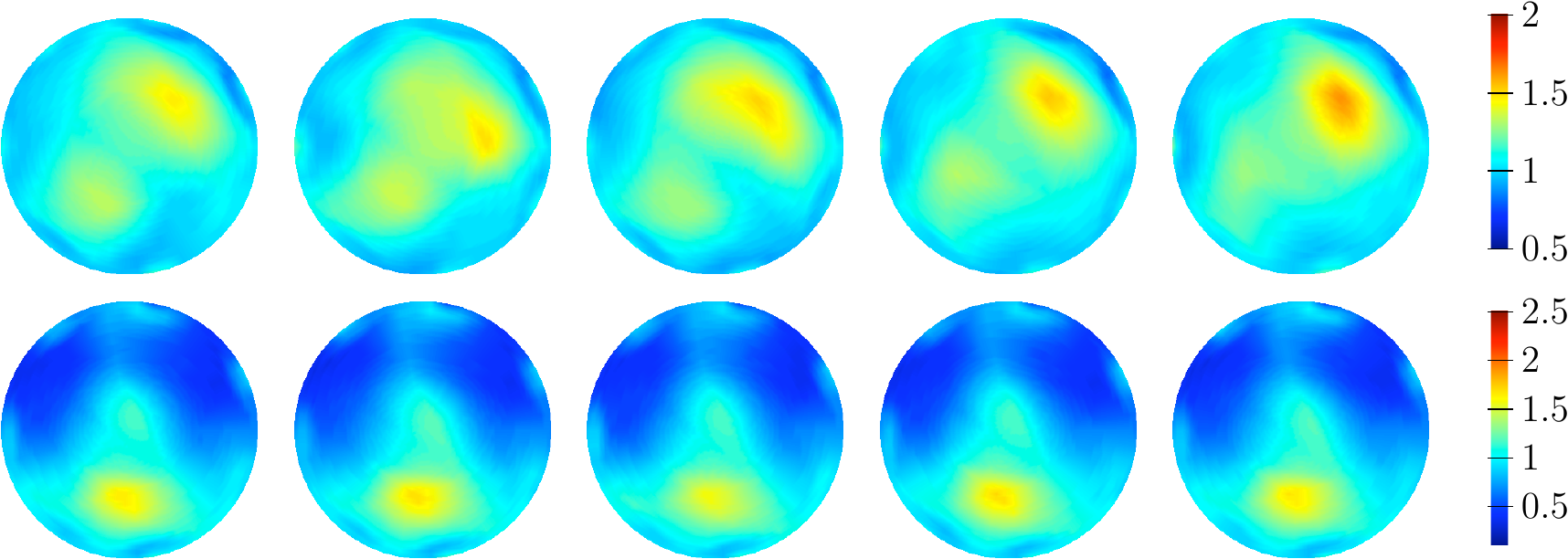}
\end{center}
\vspace{-0.1in} \caption[]{\label{fig:rols7rel} \renewcommand{\baselinestretch}{1}
\small\normalsize Realizations of $\sM(\bx)$ for $n=7$ boundary
points, $\ell = 1\%$ additive noise (\ref{eq:addN}) and regularization
parameter $\alpha=10^{-3}$ in (\ref{eq:rols}).}
\end{figure}

Here we consider the additive noise model (\ref{eq:addN}), with noise
levels $\ell = 0.01\%$ and $\ell = 1\%$. These are the levels used in
\cite{knudsen}, and we use them to compare our results with those in
\cite{knudsen}. Because solving (\ref{eq:MAP2}) with only the
upper/lower bound prior (\ref{eq:ulpr}) does not give reliable
estimates, we also use the prior (\ref{eq:gppr}). By
Remark~\ref{rem:exact}, this is equivalent to minimizing
(\ref{eq:rols}), which is computationally cheaper.

Figure \ref{fig:rols11} shows (a) the mean $\left< \s(\bx) \right>$,
(b) $\mbox{Bias}[\s(\bx)]$ and (c) the relative standard deviation
$(\mbox{Var}[\s(\bx)])^{1/2}/\left< \s(\bx) \right>$, for the noise
level $\ell = 0.01\%$.  The bias is computed with respect to
$\s_\star$, the solution of the optimization problem (\ref{eq:MAP2}),
with noiseless data, no regularization ($\alpha=0$), and a resistor
network parametrization with $n=11$ boundary nodes. The regularization
parameter $\alpha=10^{-6}$ is chosen so that both the bias and the
relative standard deviation are small.  The choice of the
regularization parameter is discussed in more detail in
section~\ref{sect:lcurve}.  We do not show realizations of the MAP
estimates, because they are close to the mean, as the standard
deviation of the reconstructions is below $10^{-3}$.

For the higher noise level $\ell =1\%$, we present in Figure
\ref{fig:rols7} the results for a smaller network, with $n = 7$
boundary nodes. Again we choose the regularization parameter $\alpha =
10^{-3}$ in such a way that both the bias and standard deviation are
small. The relative standard deviation is less than 10\%, and the
realizations of $\sM(\bx)$ shown in Figure \ref{fig:rols7rel}
resemble the mean in Figure \ref{fig:rols7}. These realizations are
comparable to the reconstructions in \cite{knudsen}.  The
reconstructions with $n=11$, noise level $\ell = 1\%$ and an
appropriate choice of the regularization parameter are qualitatively
similar to those with $n=7$ shown in Figure \ref{fig:rols7}, and thus
are not included here.

\subsubsection{\textbf{Resistor network parametrization compared to other
parametrizations}}
\label{sect:lcurve}

We now study the interplay between regularization and parametrization.
We solve the optimization problem (\ref{eq:MAP2}) with the three
parametrizations of section~\ref{sect:param}. At the noise levels
considered here, the same as in section \ref{sect:gn}, we need
regularization to get reliable estimates of the conductivity. As was
the case in section \ref{sect:gn}, the reconstructions with the
resistor network parametrization are regularized with the prior
(\ref{eq:gppr}). For the piecewise linear parametrizations, we
regularize with the Gaussian prior (\ref{eq:gcpr}), with reference
conductivity $\s_\reff \equiv 1$. Moreover, we stopped the iterations
when either the maximum number of iterations (70) is reached, or when
the norm of the gradient of the objective function at the current
iterate is smaller by a factor of $10^{-4}$ than that at the initial
iterate. We also added $10^{-8}/\C_{1,1}$ to the diagonal of the
Gauss-Newton approximation to the Hessian. Recall that $\C$ is the
covariance of the measurements.

We use two metrics to evaluate the reconstructions using different
parametrizations. The first one is the $L^2$ norm of the true bias
\[
 \mathrm{TrueBias}[\sigma] = \left(\int_{\Omega}
 (\s_\star(\bx) - \left< \s(\bx)\right>)^2 d\bx\right)^{1/2},
\] 
as a percent of the true conductivity $\s_\star$. This measures the
{\em fidelity} (in average) of our reconstructions. The second metric
is the $L^2$ norm of the standard deviation relative to $L^2$ norm of
the mean of the reconstructions
\[
 \mathrm{RelStd}[\sigma] = 
 \frac{
  \left(\int_{\Omega} \mbox{Var}[\s(\bx)] d\bx \right)^{1/2}
 }{
  \left(\int_{\Omega} \left< \s(\bx) \right>^2 d\bx \right)^{1/2}
 },
\]
which measures the {\em stability} of our reconstructions.
\begin{figure}[tb]
\begin{center}
 \begin{tabular}{c@{\hspace{1em}}c@{\hspace{2em}}c}
  & smooth & pcws const.\\[-0.1em]
  \rlab{$n=7, \ell = 1\%$} &
  \includegraphics[width=0.30\textwidth]{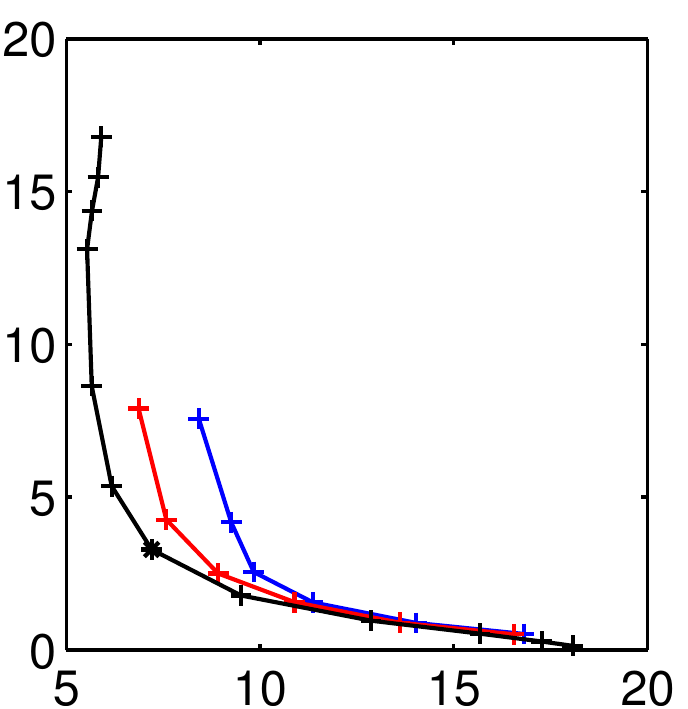} &
  \includegraphics[width=0.30\textwidth]{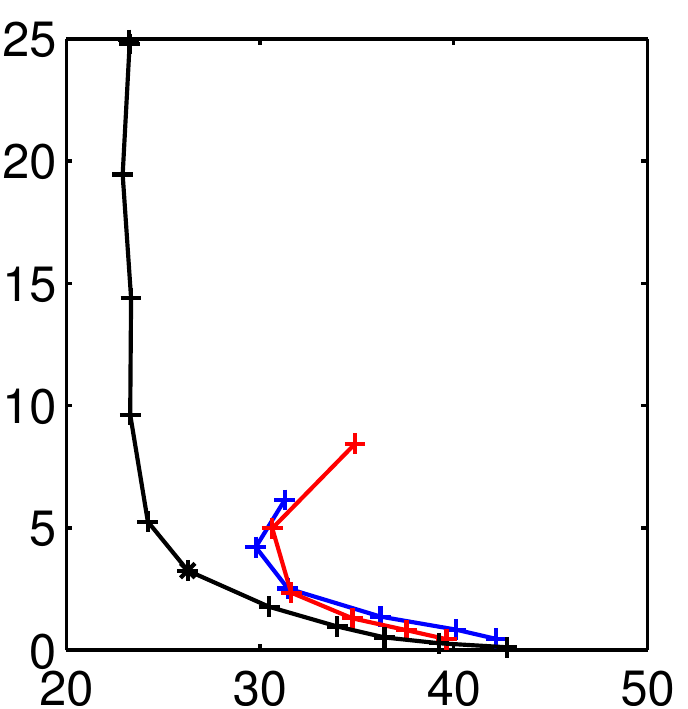}\\[-0.1em]
  \rlab{$\hspace{-0.2in} n=11, \ell = 0.01\%$} &
  \includegraphics[width=0.30\textwidth]{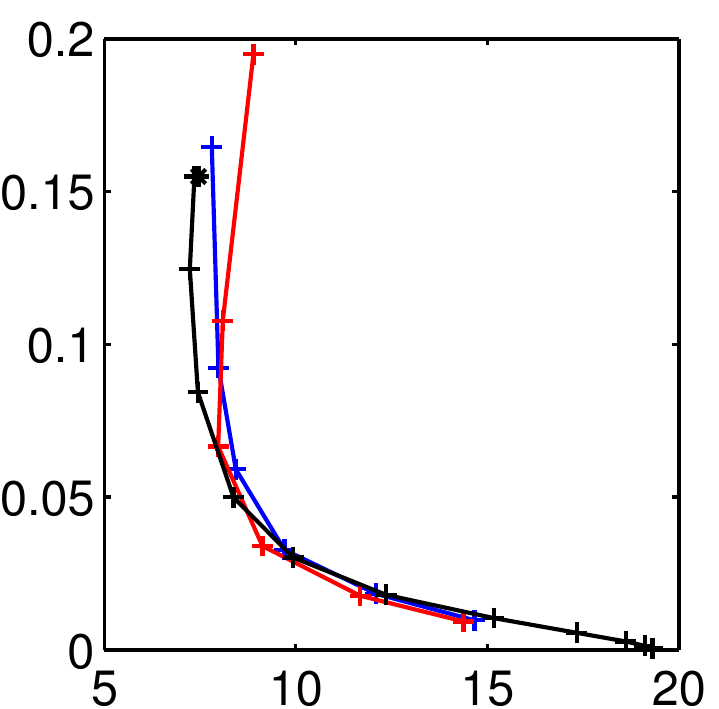} &
  \includegraphics[width=0.30\textwidth]{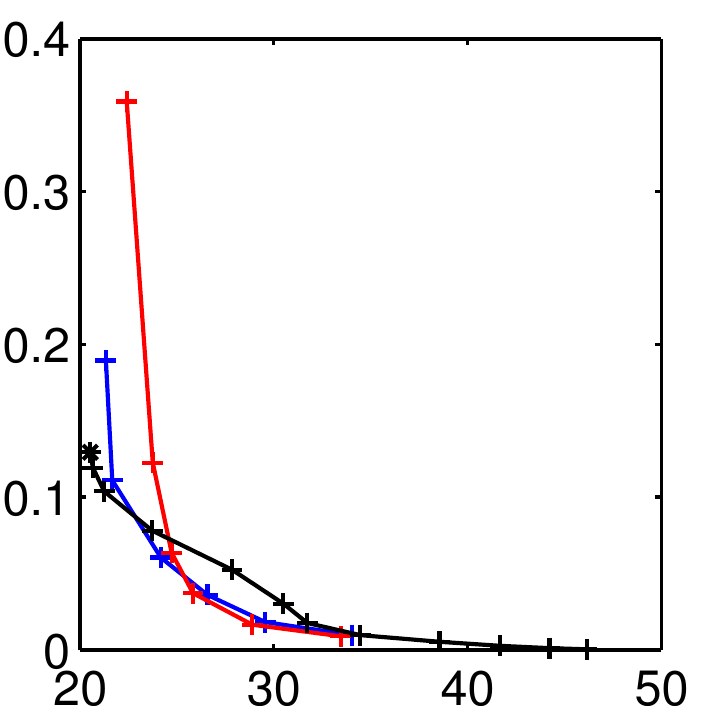}
 \end{tabular}
\end{center}
\vspace{-0.1in} \caption[]{\label{fig:lcurve} \renewcommand{\baselinestretch}{1}
\small\normalsize The true bias (in abscissa and in percent) against
the relative standard deviation (in ordinate and in percent) for
different values of the regularization parameter. The curves
correspond to the resistor network approach (black), the linear
interpolation on the optimal grid (blue) and the uniform grid
(red). The number of realizations $N=100$.  The regularization
parameter was $\alpha= 10^{-j/2}$, with $j=1,2,\ldots,6$ for the
linear interpolation (red and blue) and with $j=1,2,\ldots,12$ for the
network approach (black). The regularization parameter used for the
reconstructions in figures~\ref{fig:rols11} and \ref{fig:rols7} is
indicated with a black star.}
\end{figure}

We report in Figure~\ref{fig:lcurve}, for different values of the
regularization parameter $\alpha$, the true bias versus the relative
standard deviation.  Note the typical L-curve shape, which reflects
the trade-off between accuracy (small bias) and stability (small
standard deviation). When the regularization is not sufficient, the
bias is small but the standard deviation is large (vertical branch).
When the problem is over regularized, the bias is large and the
standard deviation is small (horizontal branch). The ``best'' choice
of the regularization parameter would be near the ``corner'' of the L
shape.

The first row in Figure \ref{fig:lcurve} shows that for noise level
$\ell=1\%$ and for $n=7$ boundary nodes, the resistor network
parametrization outperforms the piecewise linear parametrizations: for
a fixed standard deviation, the bias is smaller. This is specially
noticeable for the piecewise constant conductivity.  Interestingly,
the uniform grid is slightly better than the optimal grid in this
case.  For $n=11$ boundary nodes and $\ell=0.01\%$ (second row), all
approaches give comparable results, with the resistor network 
giving bias smaller by a few percent, specially if we allow a
standard deviation above $0.1\%$.

\begin{remark}
We emphasize that the cost of solving (\ref{eq:MAP2}) with the
resistor network approach (i.e. solving (\ref{eq:rols})) is negligible
compared to computing the Jacobian $D_\Bs\Sc(\Bs)$.  Thus, the
resistor network approach takes about the same time as one step of
Gauss-Newton to solve (\ref{eq:MAP2}) with the piecewise linear
parametrizations. In the computations for Figure~\ref{fig:lcurve}, the
mean number of iterations for these linear parametrizations was at
least 8, and varied depending on the regularization parameter, the
grid and the conductivity. Therefore, the resistor network method is
at least 8 times faster than the one using conventional
discretization.
\end{remark}

\section{Summary}
\label{sect:Summary}
We presented a numerical study of the effects of noise on resistor
based inversion algorithms. The algorithms were
introduced in \cite{BorDruGue, GuevaraPhD,BDM-10,BDMG-10,MamonovPhD},
and are briefly reviewed here. We have three measurements
setups. The first assumes that the entire boundary $\B$ of the domain
$\Omega$ is accessible. The other two are for partial boundary
measurements confined to the accessible boundary $\B_A \subset
\B$. One setup assumes one sided measurements, with $\B_A$ consisting
of a segment of $\B$. The inversion algorithm is introduced in
\cite{BDMG-10,MamonovPhD}. The other setup is two sided, with $\B_A$
consisting of two disjoint segments of $\B$. 
The inversion amounts to defining a reconstruction mapping $\Sc(\Bs)$,
that takes a vector $\Bs=\bg/\bg^{(1)}$ of ratios of positive
conductances of a network $(\Gamma,\bg)$ and a reference network
$(\Gamma,\bg^{(1)})$, to continuous conductivity functions defined in
$\Omega$. The network has a special graph $\Gamma$ that is adapted to
the measurement setup and which allows the conductors $\bg$ to be
determined uniquely from measurements of the DtN map. The 
mapping $\Sc(\Bs)$ involves a Gauss-Newton iteration that minimizes a
preconditioned data misfit in the least-squares sense. 

Our study considers three different parametrizations of the unknown
conductivity with $g$ degrees of freedom. The first two are piecewise
linear interpolations on an equidistant grid and on the optimal grid,
respectively. The third parametrization is based on resistor networks. 

For the linearized problem, the piecewise linear parametrizations give
large variances of the MAP estimates, even if we use the optimal grids.
The resistor network parametrization is superior because the variances
of the MAP estimates are lower and do not increase toward the
inaccessible part of the domain.

The statistical study of the non-linear problem shows that when no
additional prior (regularization) is introduced, and the noise is very
small, using the resistor network parametrization gives reconstructions
with small bias and the variance of the MAP estimates
is very close to the optimal Cram\'er-Rao
bound.  For larger noise, we regularize the resistor based inversion
with a prior on the conductances, and we compare the results with those
of output least squares with piecewise linear parametrizations of the
conductivity, on uniform and optimal grids and regularized with a
Gaussian prior on the conductivity.  The study assumes realistic noise
levels \cite{knudsen}. All three parametrizations give a trade-off
between accuracy (small bias) and statistical stability (small standard
deviation of the estimates).  However, the resistor based
parametrization consistently outperforms the piecewise linear ones,
giving a smaller bias for a fixed standard deviation in the
reconstructions. The quality of the reconstructions is
comparable to that in \cite{knudsen}.

Our regularization priors are very simple.  If additional prior
information is available, the results of the resistor based inversion
can be greatly improved, by enlarging the space of the Gauss-Newton
iterates, beyond the span of the sensitivity functions
of the reconstruction mapping. This was shown e.g. in \cite[\S
7.1]{BorDruGue}.

From the computational point of view, the inversion with resistor
networks can be done at roughly the cost of one Gauss-Newton iteration
for a conventional output least squares method, with the same number
of degrees of freedom of the parametrization.  In our numerical
experiments, the resistor network inversion was at least eight times 
faster.

\section*{Acknowledgements}
The work of L. Borcea was partially supported by the National Science
Foundation grants DMS-0934594, DMS-0907746 and by the Office of Naval
Research grant N000140910290. The work of F. Guevara Vasquez was
partially supported by the National Science Foundation grant
DMS-0934664. The work of A. Mamonov was partially supported by the
National Science Foundation grants DMS-0914465 and DMS-0914840.  The
authors were also partially supported by the National Science
Foundation and the National Security Agency, during the Fall 2010
special semester on inverse problems at MSRI, Berkeley. We are
grateful to Vladimir Druskin for sharing his deep insight of optimal
grids.

\appendix 
\section{Prior distributions}
\label{app:priors}
\begin{enumerate}
\item[(1)] {\bf Upper/lower bound prior}: We use this prior alone to
  explore the effect of the parametrization on the stability of the
  reconstructions, at small noise levels.  It states that the
  conductivity is positive and bounded.  Let $ \mathbb{S} = \{
  \Bs \in \R^g: ~ ~ [\Sc(\Bs)](\bx) \in (0,\s_\mathrm{max}), \quad \bx
  \in \bar{\Omega} \}  $ be the set
  of parameters mapped by $\Sc$ to positive conductivity functions
  bounded by $\s_\mathrm{max}$ in $\bar{\Omega}$.  The prior is
\begin{equation}
\ppr^{(UL)}(\Bs) = \frac{1_{\mathbb{S}}(\Bs)}{|\mathbb{S}|},
\label{eq:ulpr}
\end{equation}
where $1_{\mathbb{S}}(\Bs)$ is the indicator function that takes value
one when $\Bs \in \mathbb{S}$, and zero otherwise, and $|\mathbb{S}|$
is the volume of $\mathbb{S}$.  When we study maximum a posterior
estimates of the conductivity in section \ref{sect:Bayespdf}, we set
$\s_{\mathrm{max}}$ to a large enough value, and keep at the same time
the number $g$ of parameters low enough, for the constraint
$[\Sc(\Bs)](\bx) \le \s_{\mathrm{max}}$ to be automatically
satisfied. However, we do enforce the positivity.

\item[(2)] {\bf Gaussian prior on the conductivity}:
  This is a Tikhonov regularization prior that is
    useful at higher noise levels \cite[Chapter
      3]{kaipio2005statistical}. It says that in addition to the
    conductivity being positive and bounded, we assume that $\s$ has a
    normal distribution with mean $\s_\reff$. The fluctuations
    $\Sc(\Bs)-\s_\reff$ are uncorrelated from point to point, and the
    pointwise variance is $\alpha^{-1}$. The prior is defined by
\begin{equation}
 \ppr^{(GC)}(\Bs) \sim \ppr^{(UL)}(\Bs) \exp [-(\alpha/2) \| \Sc(\Bs)
   -\s_\reff\|_{L^2(\Omega)}^2 ],
 \label{eq:gcpr}
\end{equation}
where the symbol ``$\sim$'' means equality up to a positive, 
multiplicative constant.
\item[(3)] {\bf Prior on the parameters}: This is also a
  Tikhonov type regularization prior that is useful at higher noise
  levels. It says that in addition to the conductivity being positive
  and bounded, the vector of the logarithm of the parameters $\Bs$ is
  normally distributed, with mean $\log(\Bs_\reff)$ and covariance
  $\alpha^{-1}I$, where $I$ is the $g \times g$ identity matrix,
\begin{equation}
 \ppr^{(GP)}(\Bs) \sim \ppr^{(UL)}(\Bs) \exp [ -(\alpha/2) \|
   \log(\Bs) - \log(\Bs_\reff) \|_2^2 ].
 \label{eq:gppr}
\end{equation}

\end{enumerate}

\section{Resistor network topologies}
\label{app:nets}

\begin{figure}[t!]
 \begin{center}
   \includegraphics[width=0.25\textwidth]{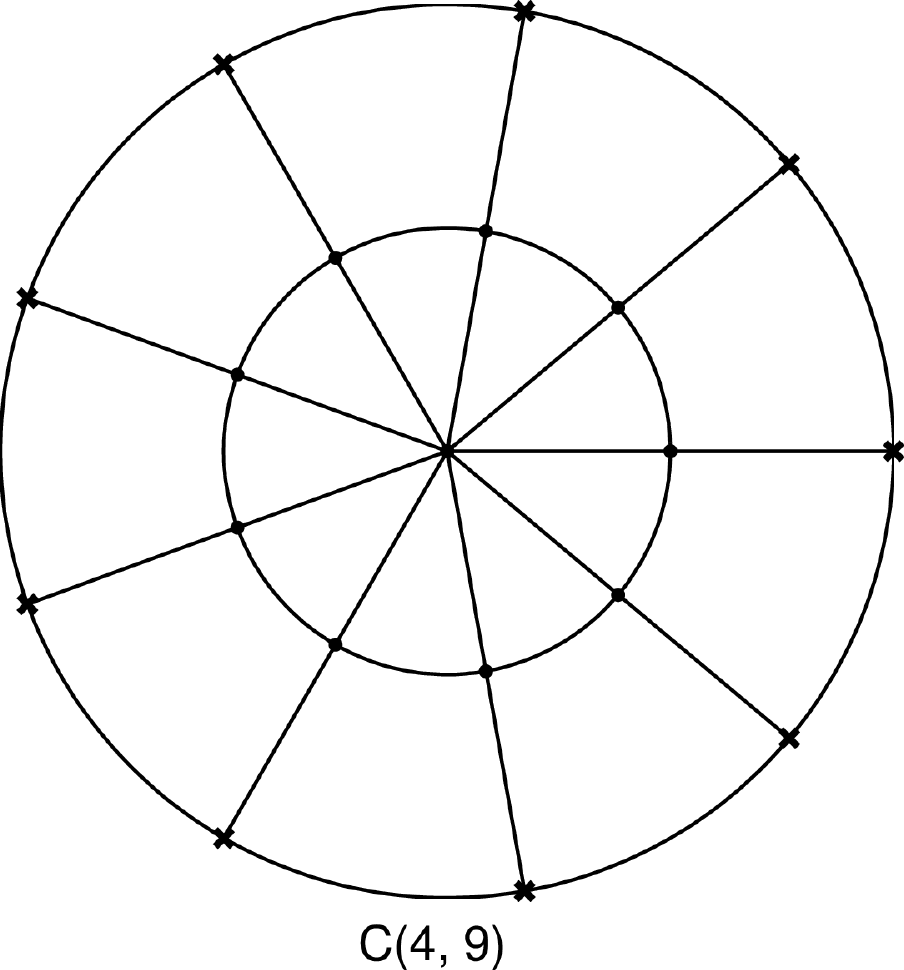}
   \hskip0.1\textwidth
   \includegraphics[width=0.25\textwidth]{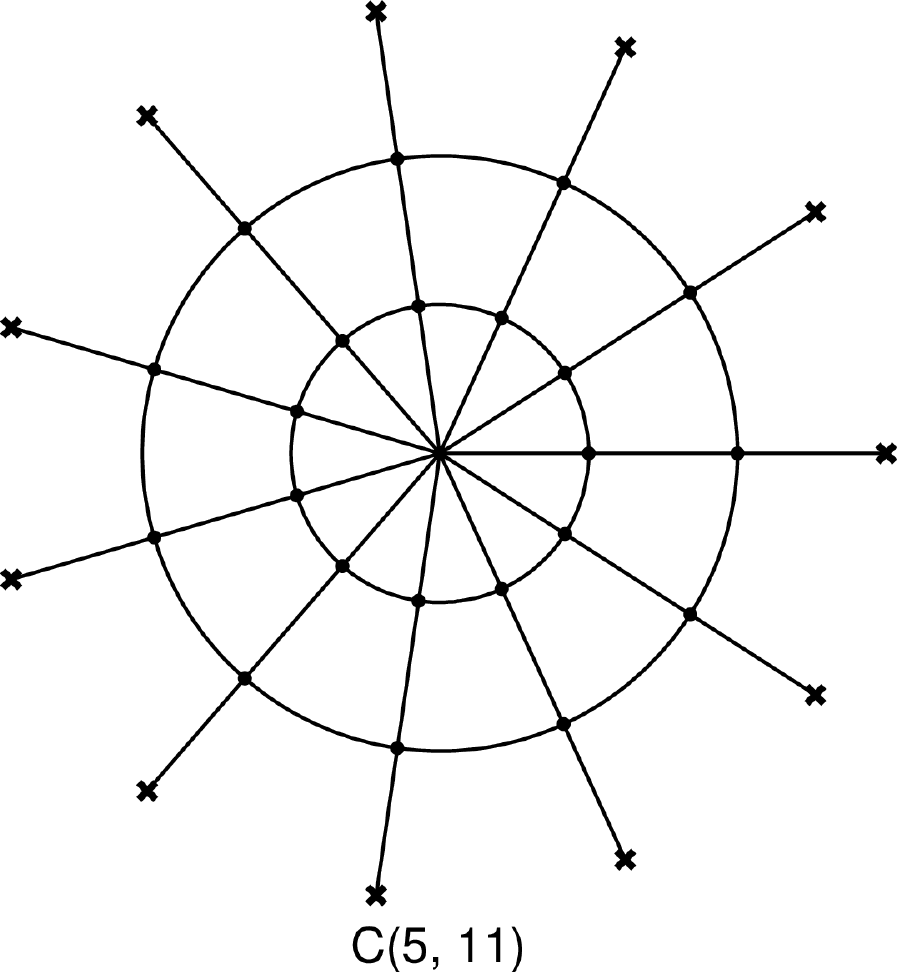}
 \end{center}
\vspace{-0.1in} \caption[Circular resistor networks]{
   \label{fig:circnet}
   \renewcommand{\baselinestretch}{1} \small\normalsize Circular
   resistor networks $C(l, n)$ with critical graphs: $l=(n-1)/2$.  The
   interior nodes are indicated with dots and the boundary nodes with
   crosses.}
\end{figure}

Resistor networks $(\Gamma,\bg)$ with \emph{circular} graphs $\Gamma =
C(l,n)$ are natural reduced models of the problem with full boundary
measurements. The notation $C(l,n)$ \cite{CurtIngMor,curtMorBook}
indicates that the graph has $l$ layers, and $n$ edges in each layer.
The edges may be in the radial direction, or transversal to it, as
illustrated in Figure \ref{fig:circnet}.  For the network to be
critical, and thus uniquely determined by $\bbF(\bg) = \bF(\s)$, we
must nave $n$ odd and $l = (n-1)/2$ \cite[Proposition 2.3, Corollary
9.4]{curtMorBook}, \cite[Theorem 2]{BorDruGue}. 

\begin{figure}[t]
 \begin{center}
 \begin{tabular}{c@{\hspace{3em}}c}
 \includegraphics[height=0.25\textwidth]{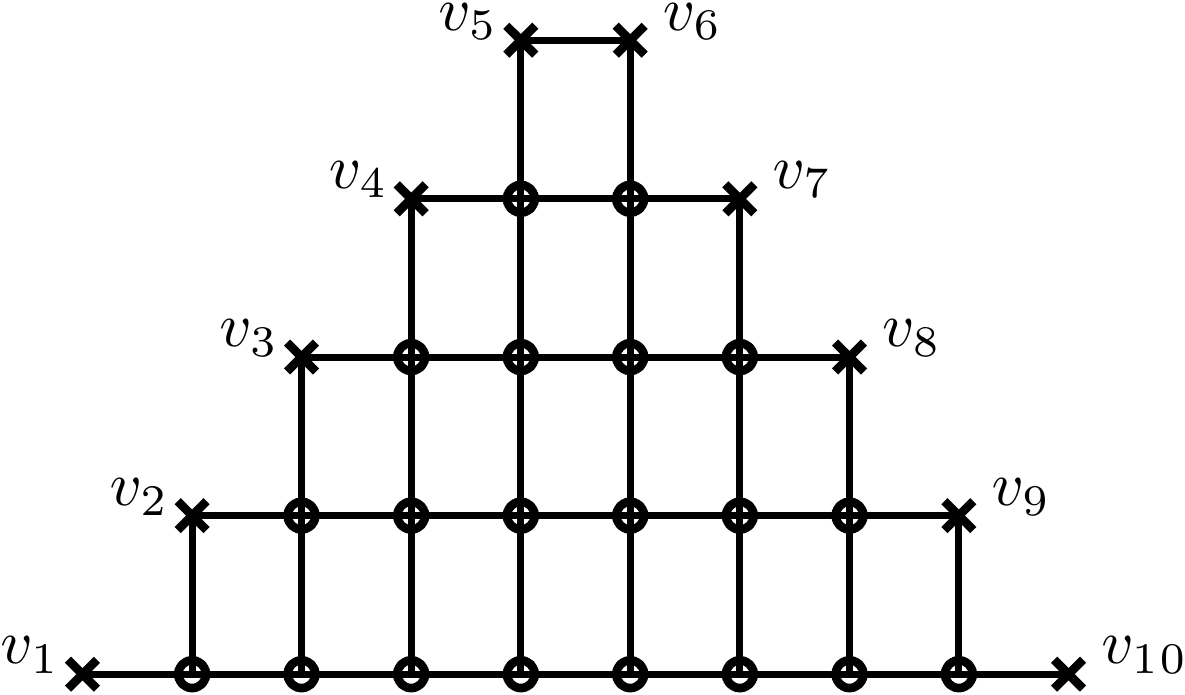}
  &
 \includegraphics[height=0.30\textwidth]{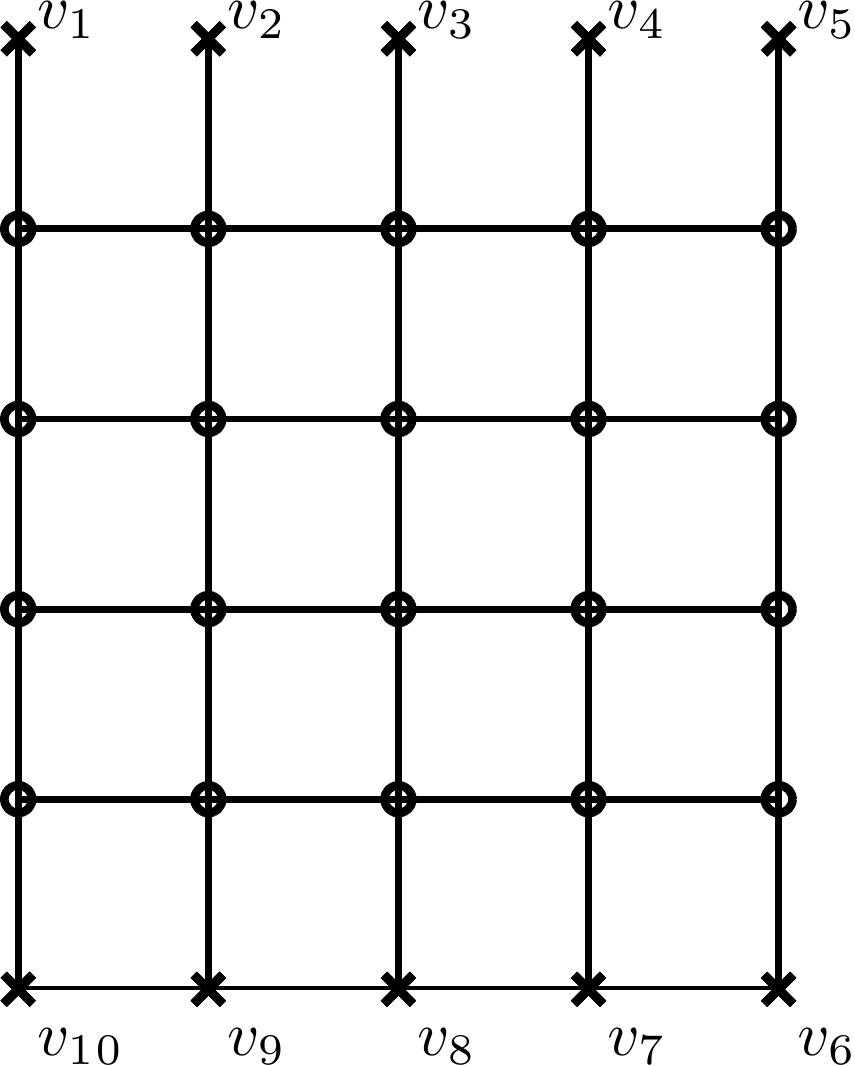} 
 \end{tabular}
 \end{center}
 \vspace{-0.1in} \caption[Networks used for partial
   measurements]{\label{fig:partialnets}
   \renewcommand{\baselinestretch}{1} \small\normalsize Resistor
   networks used for partial boundary measurements.  Left: pyramidal
   network $\Gamma_n$. Right: two-sided network $T_n$.  The boundary
   nodes $v_j$, $j=1,\ldots,n$, $n=10$ are numbered in circular order
   and are indicated by $\times$.  The interior nodes are indicated
   with $\circ$. }
\end{figure}

For the one-sided partial boundary measurements we use a different
network topology. While conformal or extremal quasiconformal
coordinate transformations allow for circular networks to be used in
the partial measurements case \cite{BDM-10}, the networks with
\emph{pyramidal} graphs $\Gamma = \Gamma_n$ are more natural
\cite{BDMG-10}. They are shown in Figure \ref{fig:partialnets} (left).
The pyramidal networks are critical and thus uniquely recoverable
\cite{BDMG-10}.  They are natural to use with one-sided partial
measurements because the sides of the pyramid, where the boundary
nodes lie, can be mapped to the accessible segment $\B_A$ of the
boundary. The base of the pyramid consists of interior nodes. They
model the lack of penetration of the currents in the part of the
domain near $\B_I$, the inaccessible boundary.  

The inversion in the two-sided case is based on \emph{two-sided}
resistor networks \cite{BDGM-11} with graph denoted by $\Gamma = T_n$,
and $n=2m$ boundary nodes. There are $m$ nodes on each segment of the
accessible boundary separated by the leftmost and rightmost interior
nodes, as illustrated in the right plot in Figure
\ref{fig:partialnets}.  These interior nodes model the lack of
penetration of the currents in the parts of the domain close to the
inaccessible boundary. The two-sided network is critical and thus can
be recovered with e.g. layer peeling \cite{BDGM-11}.
 
\section{The norm used in the noise model}
\label{app:norm}

Since the boundary $\B$ is the unit circle, we associate it with the
angle interval $[0,2\pi]$.  Consider the Fourier series operator
${\mathcal U}: \ell^2 \to L^2[0,2\pi]$, defined by
\begin{equation}
( {\mathcal U} \widehat f) (\theta) = \frac{1}{\sqrt{2 \pi}} 
\sum_{k = -\infty}^\infty \widehat f(k) e^{i k \theta},
~\mbox{and its adjoint}~
\left( {\mathcal U}^\star f\right) (k) = \frac{1}{\sqrt{2 \pi}} \int_0^{2 \pi} 
d \theta \, f(\theta) e^{-i k \theta}.
\end{equation}
The fractional Sobolev norm $H^s$ of $f$ can be written as a weighted
$\ell^2$ norm $\| ~ \|_2$, 
\begin{equation}
\| f \|_{H^s} = \| W^s {\mathcal U}^\star f \|_2,
~
\mbox{where}
~
\left( W^s \widehat v\right)_k = \left(1 + k^2\right)^{s/2}\widehat v(k), \qquad 
k \in \mathbb{Z}.
\end{equation} 
The operator norm of a linear operator $A:H^{1/2}(\B) \to H^{-1/2}(\B)$
is
\begin{equation}
\|A\|_{H^{1/2}(\B) \to H^{-1/2}(\B)} = \sup_{f \ne 0,
  f \in H^{1/2}} \frac{\| W^{-1/2} {\mathcal U}^\star f \|_2}{\|
  W^{1/2} {\mathcal U}^\star f \|_2} = \sup_{g \ne 0, g \in \ell^2}
  \frac{\| W^{-1/2} {\mathcal U}^\star A {\mathcal U} W^{-1/2} g
  \|_2}{\| g \|_2}.
\label{eq:ON}
\end{equation}
In particular, when $A = \Lambda_1$ we have $\Lambda_1 = {\mathcal U}
K {\mathcal U}^\star$, where $K \widehat v(k) = |k| \widehat v(k)$. Thus
$\|\Lambda_1\|_{H^{1/2} \to H^{-1/2}} = 1.$

We approximate the operator norm (\ref{eq:ON}) by the norm $\| ~ \|$,
as follows. In an abuse of notation, let $\Lambda_\s$ and $\Lambda_1$
be the restrictions of the continuum DtN maps to the $N$ uniformly
distributed fine grid points on $\B$. Consider the spectral 
decomposition 
\begin{equation}
\frac{2 \pi}{N} \Lambda_1 = U \Sigma U^\star, \qquad U^\star U = I.
\end{equation} 
The approximate norm is given by
 \begin{equation}
\| \Lambda_\s\| = \sup_{g \ne 0, g \in \mathbb{R}^N} \frac{2 \pi}{N}
\frac{\| (I + \Sigma^2)^{-1/4} U^\star \Lambda_\s U (I + \Sigma^2)^{-1/4} g 
  \|_2}{\| g \|_2},
\end{equation} 
which is equivalent to finding the largest eigenvalue in magnitude of
the matrix appearing in the numerator above. By construction, we have
$\| \Lambda_1\| = 1$.

\bibliography{sources} \bibliographystyle{plain}
\end{document}